%% file: paper_arxiv.tex
\documentclass[12pt]{article}
\include{_preamble}

\usepackage{arxiv}
\usepackage[inline,shortlabels]{enumitem}
\newtheorem{theorem}{Theorem}
\newtheorem{lemma}{Lemma}
\pdfoutput=1
\zexternaldocument*[supp:]{sm}

\title{\titlepaper}
\author{\authorlistarxiv}
\date{\today}

\begin{document}
\maketitle


\begin{abstract}
Causal mediation analysis has historically been limited in two important ways:
(i) a focus has traditionally been placed on binary exposures and static
interventions, and (ii) direct and indirect effect decompositions have been
pursued that are only identifiable in the absence of intermediate confounders
affected by exposure. We present a theoretical study of an (in)direct effect
decomposition of the population intervention effect, defined by stochastic
interventions jointly applied to the exposure and mediators. In contrast to
existing proposals, our causal effects can be evaluated regardless of whether
an exposure is categorical or continuous and remain well-defined even in the
presence of intermediate confounders affected by exposure. Our (in)direct
effects are identifiable without a restrictive assumption on cross-world
counterfactual independencies, allowing for substantive conclusions drawn from
them to be validated in randomized controlled trials. Beyond the novel effects
introduced, we provide a careful study of nonparametric efficiency theory
relevant for the construction of flexible, multiply robust estimators of our
(in)direct effects, while avoiding undue restrictions induced by assuming
parametric models of nuisance parameter functionals. To complement our
nonparametric estimation strategy, we introduce inferential techniques for
constructing confidence intervals and hypothesis tests, and discuss open source
software, the \texttt{medshift} \texttt{R} package, implementing the proposed
methodology. Application of our (in)direct effects and their nonparametric
estimators is illustrated using data from a comparative effectiveness trial
examining the direct and indirect effects of pharmacological therapeutics on
relapse to opioid use disorder.
\end{abstract}

\section{Introduction}\label{sec:intro}

In myriad applications, one is often interested in the effect of an exposure on
an outcome only through a particular pathway between the two. Indeed, efforts in
defining and identifying such \textit{path-specific} effects have come to
constitute a rich history in not only philosophy but also in the sciences of
statistics, causal inference, epidemiology, economics, and psychology. In each
of these disciplines, and in many others among the biomedical and social
sciences, developing a mechanistic understanding of the complexities that admit
representations as path-specific effects remains central; examples include
elucidating the biological mechanism by which a vaccine reduces infection
risk~\citep[e.g.,][]{hejazi2020efficient}, assessing the effect on preterm birth
of maternal exposure to environmental toxins, and ascertaining the effect of
novel pharmacological therapies on substance abuse disorder relapse.

The latter serves as our motivating example as we consider how exposure to
a buprenorphine dose schedule characterized by successive increases toward
a maximum dose early in treatment (versus static dose) affects the risk of
relapse to opioid use disorder, both directly and indirectly through mediating
factors such as depression and pain. Developing a detailed mechanistic
understanding of the process by which such therapeutics modulate intermediary
states is necessarily a \textit{causal} question --- one central to designing
and successively improving upon available therapies in a manner targeted towards
the mitigation of the risk of substance abuse relapse. In comparative
effectiveness trials of promising opioid use disorder therapeutics, detailed
dissections of the complex neurological and psychiatric pathways involved in the
development of addiction disorders is of clinical
interest~\citep{lee2018comparative, rudolph2020explaining}. The ability to
define and evaluate causal effects along paths involving or avoiding mediating
neuropsychiatric sequela would facilitate drug efficacy assessments; moreover,
the ability to  refine scientific conclusions based on statistical evidence
through randomized controlled trials remains integral to furthering clinical
progress.

To carefully study complex mediation relationships, a wealth of techniques
rooted in statistical causal inference have been formulated. Path
analysis~\citep{wright1934method}, perhaps the earliest example of such
methodology, directly inspired the development of subsequent techniques that
leveraged parametric structural equation
models~\citep[e.g.,][]{goldberger1972structural, baron1986moderator} for
mediation analysis. More recently, the advent of modern frameworks and
formalisms for causal inference, including nonparametric structural equation
models, directed acyclic graphs, and their underlying
do-calculus~\citep{pearl1995causal, pearl2000causality}, provided the necessary
foundational tools to express causal mechanisms without reliance on more
restrictive approaches tied to parametric modeling.

In tandem with the developments of~\citet{pearl2000causality}, similar
approaches spearheaded by~\citet{robins1986new}, \citet{spirtes2000causation},
and \citet{dawid2000causal} allowed nonparametric
formulations of mediation analysis and uncovered significant limitations of the
earlier efforts focused on structural equation models~\citep{imai2010general}.
Recent applications of modern causal models have illustrated the failings of
popular parametric modeling strategies~\citep[i.e.,][]{baron1986moderator}, in
the presence of intermediate confounders of the mediator-outcome
relationship~\citep{cole2002fallibility}. Consequently, the usually implausible
assumptions that underlie such restrictive structural equation models make these
approaches of limited applicability for examining complex phenomena in
the biomedical and health sciences.

Modern approaches to causal inference have allowed for significant advances over
the methodology of traditional mediation analysis, overcoming the significant
restrictions imposed by the use of parametric structural equation modeling. For
example,~\citet{robins1992identifiability} and~\citet{pearl2001direct}, using
distinct frameworks, provided equivalent nonparametric decompositions of the
average treatment effect (for binary exposures) into the \textit{natural}
direct and indirect effects, which quantify all effects of the exposure on the
outcome through paths avoiding the mediator and all paths involving the
mediator, respectively. Such advances were not without their limitations,
however. A key assumption of the nonparametric decomposition of the average
treatment effect is the requirement of \textit{cross-world} counterfactual
independencies (i.e., independence of counterfactuals indexed by distinct
interventions). Unfortunately, such an assumption limits the scientific
relevance of the natural (in)direct effects by making them unidentifiable in
randomized trials, directly implying that corresponding scientific claims cannot
be falsified through experimentation~\citep{popper1934logic, dawid2000causal}.
Importantly, such cross-world independencies are also unsatisfied in the
presence of intermediate confounders affected by
exposure~\citep{avin2005identifiability, tchetgen2014identification}. Given that
such confounders are challenging to rule out in practice, the natural (in)direct
effects are of limited applicability in real-world data analysis. This
incompatibility motivated the recent development of a rich family of
\textit{interventional} (in)direct effects~\citep{didelez2006direct,
vanderweele2014effect, vansteelandt2017interventional, rudolph2017robust,
nguyen2019clarifying}, which utilize a flexible joint intervention strategy to
retain identifiability in the presence of such confounding. Until quite
recently, nonparametric effect decompositions and efficiency theory were
unavailable for this class of effects, though recent efforts by
\citet{diaz2020nonparametric} and \citet{benkeser2020nonparametric} resolved
this gap in the literature. Like their natural effect counterparts, the
interventional effects are limited to settings with binary exposures.
Our work outlines a general class of causal (in)direct effect estimands that do
not require the cross-world independence condition and are robust to
intermediate confounding (like the interventional effects), though our effect
definitions are capable of readily accommodating exposure variables of all
varieties, resolving a significant practical limitation of both classes of
(in)direct effects.

A related thread of the literature has considered stochastic interventions,
which generalize many intervention classes. For example, within this framework,
static interventions result in post-intervention exposures that have degenerate
distributions. \citet{stock1989nonparametric} first considered the estimation of
the total effects of stochastic interventions, while many
others~\citep[e.g.,][]{didelez2006direct, haneuse2013estimation,
young2014identification} provided careful studies that expanded the underlying
theory of stochastic interventions and demonstrated their numerous applications.
Within the population intervention models
framework~\citep{hubbard2008population}, \citet{diaz2012population} formulated
total causal effects attributable to continuous-valued exposures using
a particular class of stochastic intervention. Conveniently, these causal
effects of stochastic interventions carry an interpretation echoing that of
standard regression adjustment. For example,~\citet{haneuse2013estimation}
described modified treatment policies, which assign post-intervention
counterfactuals based on the natural value of the exposure; their methods were
demonstrated in the context of reducing surgical time for non-small-cell lung
cancer operations.
Stochastic interventions have also successfully been applied to binary
exposures: \citet{kennedy2019nonparametric} proposed incremental propensity
score interventions and demonstrated their use in longitudinal studies in order
to circumvent identifiability and estimation issues arising from positivity
violations. Building on this flexible framework,~\citet{diaz2020causal} proposed
a decomposition of the total effect of stochastic
interventions~\citep{diaz2012population} into the \textit{population
intervention (in)direct effects}, which are endowed with interpretations
analogous to that of the natural (in)direct effects.
The (in)direct effects of~\citet{diaz2020causal} do not require cross-world
counterfactual independencies, apply to exposure variables of all types, and
succeed in accommodating nonparametric estimation strategies. Consequently,
their population intervention (in)direct effects may be estimated without
restrictive assumptions and yield scientific results that can be tested through
randomization of both the exposure and mediator. Unfortunately, the results
of~\citet{diaz2020causal} suffer a serious shortcoming --- these effects cannot
be identified in the presence of mediator--outcome confounders affected by
exposure.
In this vein, our work formulates alternative (in)direct effect estimands that
retain the flexibility of the (in)direct effects of~\citet{diaz2020causal};
however, our identification strategy emphasizes effects
robust to this form of confounding, which is accomplished by leveraging joint
stochastic interventions on the exposure and mediator.

In the present work, we outline a general framework encompassing many prior
causal mediation analysis approaches, including the natural (in)direct effects,
their interventional effect counterparts, and the stochastic (in)direct effects.
Building upon the foundations laid by \citet{diaz2020causal}, the introduced
class of mediation effects originate from combining the novel lines of inquiry
established in the distinct literatures on stochastic interventions and the
interventional effects; accordingly, we denote these \textit{stochastic
interventional (in)direct effects}. Our proposed class of effects are the first
to simultaneously avoid the requirement of cross-world counterfactual
independencies; leverage stochastic interventions to be applicable to binary,
categorical, and continuous-valued exposures; and remain identifiable despite
intermediate confounding. Our contributions apply to a broader class of
exposures than the interventional
effects~\citep[e.g.,][]{diaz2020nonparametric, benkeser2020nonparametric} while
generalizing stochastic (in)direct effects~\citep[i.e.,][]{diaz2020causal} to
accommodate the presence of intermediate confounders. While our robust and
flexible causal mediation analysis framework subsumes prior classes of effect
definitions, this is far from enough for the successful application of our
proposed (in)direct effects. To this end, we develop novel efficiency theory and
efficient nonparametric estimators of this broad class of causal mediation
parameters, within the frameworks of one-step~\citep{pfanzagl1985contributions,
bickel1993efficient} and targeted minimum loss
estimation~\citep{vdl2006targeted, vdl2011targeted}. These flexible estimators
have desirable asymptotic properties even when nuisance parameter functionals
are estimated via machine learning; moreover, they are endowed with a form of
multiple robustness producing consistent point estimates under several
configurations of nuisance parameter misspecification. Lastly, we provide
implementations of our methodological advances in our free and open source
\texttt{medshift}~\citep{hejazi2020medshift} package, for the \texttt{R}
language and environment for statistical computing~\citep{R}.

\section{Mediation analysis for the population intervention
  effects}\label{sec:backg}
\begin{figure}[!htb]
  \centering
  \begin{tikzpicture}
    \Vertex{0, -1}{L}
    \Vertex{-4, 0}{W}
    \Vertex{0, 0}{Z}
    \Vertex{-2, 0}{A}
    \Vertex{2, 0}{Y}
    \ArrowR{W}{L}{black}
    \Arrow{L}{Z}{black}
    \Arrow{W}{A}{black}
    \Arrow{A}{Z}{black}
    \Arrow{Z}{Y}{black}
    \Arrow{A}{L}{black}
    \Arrow{L}{Y}{black}
    \ArrowL{W}{Y}{black}
    \ArrowL{A}{Y}{black}
    \ArrowL{W}{Z}{black}
  \end{tikzpicture}
  \caption{Directed Acyclic Graph of NPSEM (\ref{eq:npsem}).}
  \label{fig:dag}
\end{figure}
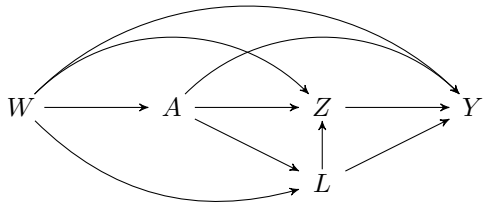

Let $A$ denote a continuous or categorical exposure, $Y$ denote a continuous or
binary outcome, $Z$ denote mediator(s), $W$ denote a vector of observed
pre-exposure covariates, and $L$ denote an intermediate (mediator--outcome)
confounder affected by exposure. The nonparametric structural equation model
(NPSEM) formalizes the problem:
\begin{equation}\label{eq:npsem}
  W = f_W(U_W); A = f_A(W, U_A); L = f_L(A, W, U_L);
  Z = f_Z(L, A, W, U_Z); Y = f_Y(Z, L, A, W, U_Y).
\end{equation}
In the NPSEM (\ref{eq:npsem}), $U=(U_W,U_A,U_L,U_Z,U_Y)$ is a vector of
exogenous factors, and the functions $f$ are assumed deterministic but unknown.
This mechanistic model is assumed to generate the observed data $O$; it encodes
several fundamental assumptions. First, an implicit temporal ordering $W
\rightarrow A \rightarrow L \rightarrow Z \rightarrow Y$ is assumed. Second,
each variable (i.e., $\{W, A, L, Z, Y\}$) is assumed to be generated from the
corresponding deterministic function of the observed variables that precede it
temporally, plus an exogenous variable denoted by $U$. Each exogenous variable
is assumed to contain all unobserved causes of the corresponding observed
variable. For a random variable $X$, let $X_a$ denote the counterfactual outcome
observed in a hypothetical world in which $\P(A=a)=1$. For example, we have
$L_a = f_L(a, W,U_L)$, $Z_a=f_Z(L_a, a, W,U_Z)$, and $Y_a=f_Y(Z_a, L_a, a, W,
U_Y)$. Likewise, we let $Y_{a,z} = f_Y(z, L_a, a, W,U_Y)$ denote the value of
the outcome in a hypothetical world where $\P(A = a, Z = z)=1$.
Figure~\ref{fig:dag} represents model~\eqref{eq:npsem} in terms of a directed
acyclic graph (DAG).

Letting $O = (W, A, L, Z, Y)$ represent a random variable with distribution
$\P$, we denote by $O_1, \ldots, O_n$ a sample of $n$ i.i.d.~observations of
$O$. We let $\P f = \int f(o)\dd \P(o)$ for a given function $f(o)$. We use
$\P_c$ to denote the joint distribution of $(O,U)$, and let $\E$ and $\E_c$
denote corresponding expectation operators. We use $\Pn$ to denote the empirical
distribution of $O_1, \ldots, O_n$, and assume $\P \in \M$, where $\M$ is the
nonparametric statistical model defined as all continuous densities on $O$ with
respect to a dominating measure $\nu$. Let $\p$ denote the corresponding
probability density function. We use $\g(a \mid w)$ and $\e(a \mid z, w)$ to
denote the probability density function or the probability mass function of $A$
conditional on $W = w$ and $(Z, W)$, respectively; $\m(z,l,a,w)$ to denote the
outcome regression function $\E(Y \mid Z = z,L=l,A = a,W = w)$. Let
$\g(\cdot\mid w)$ and $\e(\cdot\mid z,w)$ be dominated by a measure $\kappa(a)$
(e.g., the counting measure for binary $A$ and the Lebesgue measure for
continuous $A$). 
In constructing our estimators, we will use
\begin{equation}
  \frac{\p(z\mid w)}{\p(z\mid a, w)} = \frac{\g(a\mid w)}{\e(a\mid z,w)};
    \quad\quad \frac{\p(z\mid a,w)}{\p(z\mid l,a,w)}=
    \frac{\p(l\mid a,w)}{\p(l\mid z,a,w)}\label{eq:parameterize}
\end{equation}
as such parameterizations allow for estimation and integration with respect to
multivariate conditional densities on the mediator $Z$ to be avoided. We use
$\cal W, \cal A, \cal L, \cal Z$, and $\cal Y$ to denote the support of the
corresponding random variables.

Causal effects are defined in terms of hypothetical interventions on the
NPSEM~\eqref{eq:npsem}. In particular, consider an intervention in which the
structural equation corresponding to $A$ is removed, with the exposure drawn
instead from a user-specified distribution $g_\delta(a \mid w)$, which may
itself depend on the natural exposure distribution and a user-specified
parameter $\delta$. Going forward, we let $A_\delta$ denote a draw from
$g_\delta(a \mid w)$. Alternatively, such modifications can occasionally be
described in terms of an intervention in which the structural equation
corresponding to $A$ is removed and the exposure is set equal to a hypothetical
regime $d(A, W)$. Regime $d$ depends on the exposure level $A$ that would have
been assigned in the absence of the regime as well as on $W$. The latter
intervention has been referred to as depending on the \textit{natural value of
treatment}, or as a \textit{modified treatment
policy}~\citep[MTP;][]{haneuse2013estimation}. For such interventions,
\citet{haneuse2013estimation} introduced the assumption of \textit{piecewise
smooth invertibility}, which ensures that the change of variable formula can be
used when computing integrals over $\cal A$:
\begin{assumptioniden}[Piecewise smooth invertibility]\label{ass:inv}
  For each $w \in \cal W$, assume that the interval
  ${\cal I}(w) = (l(w,), u(w))$ may be partitioned into subintervals
  ${\cal I}_{\delta,j}(w):j = 1, \ldots, J(w)$ such that $d(a, w)$ is equal to
  some $d_j(a, w)$ in ${\cal I}_{\delta,j}(w)$ and $d_j(\cdot,w)$ has inverse
  function $h_j(\cdot, w)$ with derivative $h_j'(\cdot, w)$.
\end{assumptioniden}
Assumption~\ref{ass:inv} can be used to show that the intervention drawing
$A_{\delta}$ from the post-intervention distribution $\g_\delta(a \mid w)$ can
be interpreted on the individual level. \citet{young2014identification} provide
a discussion comparing and contrasting the interpretation and identification of
these two interventions. Such stochastic interventions can be used to define the
\textit{population intervention effect (PIE)} of $A$ on $Y$. To illustrate,
suppose $A$ to be continuous-valued and assume the distribution of $A$
conditional on $W = w$ is supported in the interval $(l(w), u(w))$. Then, one
may define
\begin{equation}\label{eq:defdshift}
  d(a, w) =
  \begin{cases}
    a - \delta & \text{if } a > l(w) + \delta \\
    a & \text{if } a \leq l(w) + \delta,
  \end{cases}
\end{equation}
where $0 < \delta < u(w)$ is an arbitrary prespecified value. We
can alternatively define a tilted intervention distribution as
\begin{equation}\label{eq:tilt}
  \g_\delta(a \mid w) = \frac{\exp(\delta a) \g(a \mid w)}
  {\int \exp(\delta a) \g(a\mid w)\dd\kappa(a)},
\end{equation}
for $\delta \in \mathbb R$. 
\citet{kennedy2019nonparametric} proposed a form of exponential
tilting~\eqref{eq:tilt} under the parameterization $\delta' = \exp(\delta)$,
appropriate for incremental interventions on the propensity score for binary
$A$.
Two key distinctions between the interventions defining modified treatment
policies~\eqref{eq:defdshift} and exponential tilting~\eqref{eq:tilt} can be
helpful in differentiating between the two in practice. Firstly, through
Assumption~\ref{ass:inv}, intervention~\eqref{eq:defdshift} defines
counterfactual exposures equipped with an individual-level interpretation,
whereas the intervention~\eqref{eq:tilt} only admits interpretation as a random
draw from a post-intervention distribution. Secondly,
intervention~\eqref{eq:defdshift} has been historically studied in settings with
continuous (or ordinal)
exposures~\citep[e.g.,][]{diaz2013assessing,hejazi2020efficient}, though it may
apply more broadly. While intervention~\eqref{eq:tilt} applies readily to
exposures of any type (e.g., continuous,
categorical),~\citet{kennedy2019nonparametric} introduced it to construct total
effects with weakened positivity requirements in longitudinal studies with
time-varying binary exposures.
\citet{diaz2020causal} provide a careful study of these interventions for
mediation analysis, introducing novel (in)direct effects and efficiency theory.
Their contributions assume the absence of intermediate confounding; our
generalization remedies this inconvenient shortcoming.

\subsection{Stochastic Mediation Effects}

\cite{diaz2020causal} defined the (in)direct effect of $A$ on $Y$ in
terms of a decomposition of the total effect of a stochastic
intervention. In particular, the total effect $\E(Y - Y_{A_\delta})$ may be
decomposed as the sum of the population intervention direct and indirect effects
(PIDE; PIIE):
\begin{align}
  \text{PIDE} &= \E_c\{f_Y(Z, L, A, W,U_Y) -
                f_Y(Z, L_{A_\delta},A_\delta,W,U_Y)\}\\ \nonumber
  \text{PIIE} &= \E_c\{f_Y(Z, L_{A_\delta},A_\delta,W,U_Y) -
                f_Y(Z_{A_\delta},L_{A_\delta},A_\delta,W,U_Y)\}.
\end{align}
Upon inspection, the definitions above reveal that the direct effect
measures the effect through paths \emph{not} involving the mediator
(i.e., $A \rightarrow Y$ and $A\rightarrow L \rightarrow Y$), whereas the
indirect effect measures the effect through paths involving the mediator
(i.e., $A\rightarrow Z \rightarrow Y$ and $A\rightarrow L \rightarrow Z
\rightarrow Y$).

Unfortunately, the population intervention (in)direct effects are not generally
identified in the presence of an intermediate confounder affected by exposure
such as in the DAG in Figure~\ref{fig:dag}~\citep{diaz2020causal}. This is due
to the dual role of $L$ as a confounder of the relation between $Z$ and $Y$,
which requires adjustment, and a variable on the path from $A$ to $Y$, which
precludes adjustment.
Note that~\citet{vansteelandt2012natural} attempted to circumvent these
restrictions in the context of direct effects, while~\citet{fulcher2019robust}
complemented their study by introducing an alternative effect decomposition that
yielded indirect effects with weaker identification requirements. In particular,
\citet{fulcher2019robust} contrive an indirect effect deifnition that is
identifiable under unmeasured baseline confounding of the exposure--outcome
relationship and that is formulated by the application of a stochastic
intervention to the mediator. By contrast, \citet{diaz2020causal} leverage joint
stochastic interventions on the exposure and mediator in their effect
decomposition, which can be made to achieve a similar identification property.
(Coincidentally, \citet{fulcher2019robust} refer to their effects as
``population intervention (in)direct effects,'' terminology also used by
\citet{diaz2020causal}, though the effects of the former are in some ways more
restrictive than those proposed by the latter.)
The interventional effects~\citep{vanderweele2014effect} resolve the
identification issue brought on by $L$, though their limitation to static
interventions acting upon binary exposures is a significant hurdle to their use.
Next, we present a solution to this complication using a joint stochastic
intervention on the exposure $A$ and mediator $Z$. We also show that the effects
defined in this manuscript are a generalization of the effects
of~\cite{diaz2020causal} in the sense that the former reduce to the latter in
the absence of intermediate confounding.

\subsection{Stochastic Interventional Mediation Effects}

To introduce (in)direct effects robust to the presence of intermediate
confounders, we draw upon ideas first outlined by~\cite{didelez2006direct}
and~\cite{vdl2008direct}, later formalized or subsumed
by~\cite{vanderweele2014effect} and~\cite{vansteelandt2017interventional}. Owing
to their definition in terms of stochastic interventions on the mediator, these
(in)direct effects have been collectively termed \emph{interventional effects}.
We leverage two types of stochastic interventions: one on the exposure $A$,
which defines the intervention of interest, and one on the mediator $Z$, which
is used to achieve identifiability of the effects. Following the convention of
the literature, we term stochastic interventions on the mediator $Z$
\textit{interventional}, while reserving the label of \textit{stochastic} to
refer only to interventions on the exposure $A$. To proceed, let $G_\delta$
denote a random draw from the distribution of $Z_{A_\delta}$ conditional on
$(A_\delta, W)$, and let $G$ denote a random draw from the distribution of $Z$
conditional on $(A, W)$.
To distinguish $G_{\delta}$ from the previously defined $g_{\delta}(a \mid w)$,
note that the latter is the post-intervention distribution of the exposure,
based on the user-specified scalar $\delta$, and gives rise to the
counterfactual $A_{\delta}$, while the former is a counterfactual arising from
a draw from the \textit{interventional distribution} of the mediator, which
breaks the dependence of $Z$ on $L$. These draws are denoted $G_{\delta}$ (when
$A_{\delta}$ is used) or $G$ (when $A$ is used).
We consider the effect defined by $\psi_\delta
= \E_c\{Y_{A,G} - Y_{A_\delta,G_\delta}\}$. Note that the effect $\psi_{\delta}$
is distinct from the effect considered by~\cite{diaz2020causal}, which may be
expressed $\E_c\{Y_{A,Z} - Y_{A_\delta,Z_\delta}\}$. The effect $\psi_\delta$
arises from fixing the mediator to a random value chosen from its distribution
among all those with a particular exposure level, rather than fixing it to what
it would have been under a particular exposure level.
The choice of intervention on $Z$ (i.e., defining $G_{\delta}$, which does not
depend on $L$) allows for the contribution of the intermediate confounder $L$
upon the mediator $Z$ to be eliminated. The reasoning behind this intervention
is as follows. Considering the DAG (see Figure~\ref{fig:dag}) corresponding to
model~\eqref{eq:npsem}, $L$ satisfies the ``recanting witness'' criterion
of~\citet{avin2005identifiability}. This introduces unidentifiability of the
natural (in)direct effects. Removal of the directed path from $L$ to $Z$ in the
interventional distribution resolves this complication, analogously to solutions
presented by those authors for the identification of simpler path-specific
effects.
Defining the effect in this way aids in achieving an identifiable decomposition
into direct and indirect effects. In particular, we may decompose this effect in
terms of stochastic interventional \textit{direct effects (DE)} and
\textit{indirect effects (IE)}:
\begin{equation}\label{eqn:pie_decomp}
  \psi(\delta) = \overbrace{\E\{Y_{A, G} -
    Y_{A_\delta, G}\}}^{\text{DE}} +
  \overbrace{\E\{Y_{A_\delta, G} -
    Y_{A_\delta, G_\delta}\}}^{\text{IE}}.
\end{equation}
Decomposition as the sum of direct and indirect effects affords an
interpretation analogous to the corresponding standard decomposition of the
average treatment effect into the natural direct and indirect
effects~\citep{pearl2001direct}. In particular, the direct effect arises
from drawing a counterfactual value of $A$ from a post-intervention
distribution while keeping the distribution of $Z$ fixed. The indirect
effect arises from replacing the distribution of $Z$ with a candidate
post-intervention distribution while holding $A$ fixed. Our proposed stochastic
interventional effects have an interpretation similar to the interventional
effects of~\citet{vanderweele2014effect}; moreover, while both effect
definitions account for the presence of an intermediate confounder, our
(in)direct effects utilize flexible, stochastic interventions on the
exposure while those of~\citet{vanderweele2014effect} are limited to static
interventions on binary exposures. By generalizing the effect definitions of
\citet{diaz2020causal}, our proposed (in)direct effects include, as special
cases, the natural (in)direct effects (under a static intervention on binary $A$
and no intermediate confounders $L$), the interventional (in)direct effects
(under a static intervention on binary $A$ and a stochastic intervention on $Z$,
allowing intermediate confounders $L$), and the stochastic (in)direct effects
(under a stochastic intervention on arbitrary-valued $A$ and no intermediate
confounders $L$).

\subsection{Identification}\label{sec:iden}

To construct estimators of our proposed causal (in)direct effects, we turn to
examining assumptions needed to estimate components of the post-intervention
quantities corresponding to counterfactuals of interest. First, we consider
Assumptions~\ref{ass:cs}--\ref{ass:ncaz}, which allow identification
of the stochastic interventional effects of Equation~\eqref{eqn:pie_decomp}:
\begin{assumptioniden}[Common support]\label{ass:cs}
  Assume $\supp\{\g_\delta(\, \cdot \mid w)\} \subseteq
  \supp\{\g(\, \cdot \mid w)\}$ for all $w\in \cal W$.
\end{assumptioniden}
\begin{assumptioniden}[Mediator positivity]\label{ass:zpos}
  Assume $\p(z \mid a, w) > 0$ and further assume $\p(z \mid l, a, w) > 0$.
\end{assumptioniden}
\begin{assumptioniden}[No unmeasured exposure-outcome confounder]\label{ass:ncay}
  Assume $Y_{a,z} \indep A \mid W$.
\end{assumptioniden}
\begin{assumptioniden}[No unmeasured mediator-outcome confounder]\label{ass:nczy}
  Assume $Y_{a,z} \indep Z \mid (L,A,W)$.
\end{assumptioniden}
\begin{assumptioniden}[No unmeasured exposure-mediator confounder]\label{ass:ncaz}
  Assume $Z_a \indep A \mid W$.
\end{assumptioniden}
Assumption~\ref{ass:ncay} states that, conditional on $W$, there is no
unmeasured confounding of the relation between $A$ and $Y$;
Assumption~\ref{ass:ncaz} states that conditional on $W$ there is no unmeasured
confounding of the relation between $A$ and $Z$; and Assumption~\ref{ass:nczy}
states that conditional on $(W,A,L)$ there is no unmeasured confounding of the
relation between $Z$ and $Y$. These assumptions are standard in causal mediation
analysis. In addition to these assumptions, standard mediation
analyses~\citep[e.g., ][]{vanderweele2014effect} require positivity assumptions
on the exposure and mediation mechanisms (e.g., Assumption~\ref{ass:zpos}). The
stochastic intervention framework we adopt does not require such assumptions, as
positivity can be arranged by definition of $\g_\delta$. For example, the
interventions in expressions (\ref{eq:defdshift}) and (\ref{eq:tilt}) satisfy
Assumption~\ref{ass:cs} by definition~\citep[see,
e.g.,][]{kennedy2019nonparametric,hejazi2020efficient,diaz2020causal}.
Notably, our effects do not require any assumption on the independence of
cross-world counterfactuals, required for identification of the natural
(in)direct effects. The cross-world independence assumption is restrictive, as
it cannot be tested under randomization, significantly limiting the scientific
relevance of these effects~\citep{diaz2020causal,popper1934logic}. Our novel
effect definitions are hardly the first to circumvent this assumption: earlier
work~\citep[e.g.,][]{didelez2006direct,vdl2008direct,
vansteelandt2012natural} provided strategies for loosening the reliance of the
natural (in)direct effects on the cross-world assumption. As with the earlier
effect definitions of~\citet{diaz2020causal}, our stochastic interventional
effects similarly eschew this restrictive condition for their identification.
Under these assumptions, the following identification results hold; proofs
appear in the \href{sm}{Supplementary Materials}.
\begin{theorem}[Identification]\label{theo:iden}
  Define
  \begin{align*}
    \theta_{1,\delta} &= \int \m(z,l,a,w)\p(l\mid a,w)\p(z\mid a,w)
        \g_\delta(a \mid w)\p(w)\dd\nu(a,z,l,w),\\
    \theta_{2,\delta} &=  \int \m(z,l,a,w)\p(l\mid a,w)\p(z\mid w)
        \g_\delta(a \mid w)\p(w)\dd\nu(a,z,l,w).
  \end{align*}
  Under Assumptions~\ref{ass:cs}--\ref{ass:ncaz}, the direct effect
  $\psid = \E\{Y_{A, G} - Y_{A_\delta, G}\}$ and indirect effect
  $\psii = \E\{Y_{A_\delta, G} - Y_{A_\delta,
  G_\delta}\}$~\eqref{eqn:pie_decomp} are identified, respectively, by
\begin{equation}
    \psid = \theta_{1,0} - \theta_{2,\delta} \quad \text{and} \quad
    \psii = \theta_{2,\delta} -\theta_{1,\delta}.
  \label{eq:defpsi}
\end{equation}
\end{theorem}
A consequence of this identification result is that the definitions reduce to
the stochastic (in)direct effects of~\cite{diaz2020causal} in the absence of
intermediate confounders $L$. Importantly, this implies that our estimators can
be safely used in the absence of intermediate confounders; furthermore, it
implies that the corresponding estimates may be interpreted in terms of
a decomposition of the population intervention effect $\E_c\{Y
- Y_{A_\delta}\}$, which, like the interventional effect $\psi_\delta
= \E_c\{Y_{A,G} - Y_{A_\delta,G_\delta}\}$, may be of scientific relevance.

Examination of Definition~\eqref{eq:defpsi} reveals that evaluation of $\psid$
and $\psii$ requires access to $\theta_{2,\delta}$, which is based on
$\g_{\delta}(a \mid w)$. The quantities $\psid$ and $\psii$ further require
access, respectively, to either $\theta_{1,0}$ or $\theta_{1, \delta}$, which
draw upon $\g_{\delta}(a \mid w)$. Notably, $\theta_{1,0}$ is
\textit{not} the population mean outcome, rather it is the ``interventional''
counterfactual mean arising from changing the distribution of the mediator from
$\p(Z \mid L, A, W)$ to $\p(Z \mid A, W)$, resulting in the counterfactuals $G$
and $G_{\delta}$. In fact, this induced independence of $L$ and $Z$, conditional
on $A$ and $W$, recovers the effects of~\citet{diaz2020causal} when there is no
post-exposure confounder $L$. We next turn our attention to developing
efficiency theory for estimation of the statistical parameter
$\theta_{j,\delta}: j = 1,2$, which depends on the observed data distribution
$\P$.

\section{Optimality theory for estimation of the direct
  effect}\label{sec:optimal}
Thus far, we have discussed the decomposition of the effect of a stochastic
intervention into direct and indirect effects and have provided identification
results under standard assumptions. Next, we develop efficiency theory for
estimating $\theta_{1,\delta}$ and $\theta_{2,\delta}$ in the nonparametric
model $\M$. To do so, we introduce the \textit{efficient influence function}
(EIF), which characterizes the asymptotic behavior of all regular and
asymptotically linear estimators~\citep{bickel1993efficient}. Three common
frameworks exist for constructing locally efficient estimators based on the EIF:
(i) estimating equations~\citep[e.g.,][]{vdl2003unified}, (ii) one-step bias
correction~\citep[e.g.,][]{pfanzagl1985contributions, bickel1993efficient}, and
targeted minimum loss estimation~\citep{vdl2006targeted, vdl2011targeted}.

As a consequence of its representation in terms of orthogonal score equations,
the EIF allows the construction of consistent estimators of the target parameter
even when certain nuisance components are inconsistently estimated.
Second-order bias terms may be derived from asymptotic analysis of estimators
constructed based on the EIF --- often, these estimators require slow
convergence rates (e.g., $n^{-1/4}$) for the nuisance parameters involved,
in order for the estimators to be regular and asymptotically linear (thereby
achieving a Gaussian limit distribution), achieve $\sqrt{n}$-consistency, and
exhibit asymptotic efficiency. Importantly, it is this rate-convergence property
that enables the use of flexible, data adaptive regression techniques in
estimating these quantities.

In Theorem~\ref{theo:eif}, we present the EIF for a general stochastic
intervention. Although the components of the EIF associated with $(W,L,Z,Y)$ are
the same, the component associated with the model for the distribution of $A$
must be computed on a case-by-case basis, that is, for each intervention of
interest. Lemmas~\ref{lemma:mtp} and~\ref{lemma:tilt} present such components
for modified treatment policies satisfying Assumption~\ref{ass:inv} and for
exponential tilting, respectively. In Theorem~\ref{theo:eif} below, we present
a representation of the EIF that circumvents the challenging computation of
multivariate integrals over $Z$. To introduce the EIF, we define the following
auxiliary nuisance parameters:
\begin{equation}
  \begin{split}
  \uu(z,a,w) & = \int \m(z,l,a,w) \dd\P(l\mid a,w);\\
  \vv(l,a,w) & = \int \m(z,l,a,w) \dd\P(z\mid a,w);\\
  \s(l,a,w) & = \int \m(z,l,a,w) \dd\P(z\mid w);
\end{split}\quad\quad
\begin{split}
\bar \uu(a,w) & = \int \uu(z,a,w) \dd\P(z\mid a,w)\\
  \bar \vv(a,w) & = \int \vv(l,a,w) \dd\P(l\mid a,w)\\
  \bar \s(a,w) & = \int \s(l,a,w) \dd\P(l\mid a,w)
\end{split}\label{eq:nuis}
\end{equation}
Proofs for the following results are detailed in the \href{sm}{Supplementary
Materials}.
\begin{theorem}[Efficient influence functions]\label{theo:eif} Define
\begin{equation*}
  H^1_{\P,\delta}(a,z,l,w) \coloneqq \frac{\g_\delta(a\mid
    w)}{\g(a\mid w)}\frac{\p(z\mid a,w)}{\p(z\mid a,l,w)};\quad
  H^2_{\P,\delta}(a,z,l,w) \coloneqq \frac{\g_\delta(a\mid
    w)}{\g(a\mid w)}\frac{\p(z\mid w)}{\p(z\mid a,l,w)}.
  \label{eq:Hs}
\end{equation*}
The efficient influence functions for $\theta_{j, \delta} : j = 1, 2$ in the
nonparametric model are equal to
$D^j_{\P, \delta}(o) - \theta_{j,\delta}$, where
$D^j_{\P,\delta}(o) = S^j_{\P,\delta}(o) + S^{j,A}_{\P,\delta}(o)$ and
  \begin{align}
    S^1_{\P,\delta}(o) & =  H^1_{\P,\delta}(a,z,l,w)\{y -
             \m(z,l,a,w)\}\label{eq:scorey1}\\
                         &+ \frac{\g_\delta(a\mid
                           w)}{\g(a\mid w)}\big[\vv(l,a,w) -
                           \bar\vv(a,w)  + \uu(z,a,w)
                           -\bar\uu(a,w)\big]\label{eq:scorelz1}\\
                         &+\int \bar\uu(a,w)\g_\delta(a\mid
                           w)\dd\kappa(a)\notag\\
    S^2_{\P,\delta}(o)  & =  H^2_{\P,\delta}(a,z,l,w)\{y -
                            \m(z,l,a,w)\}\label{eq:scorey2}\\
                       & + \frac{\g_\delta(a\mid
                            w)}{\g(a\mid w)}\{\s(l,a,w) - \bar
                         \s(a,w)\}\label{eq:scorelz2}\\
                         & +\int \uu(z,a,w)\g_\delta(a\mid
                           w)\dd\kappa(a)\notag,
  \end{align}
  and $S^{1,A}_{\P,\delta}(o)$, $S^{2,A}_{\P,\delta}(o)$ are the respective
  efficient score functions of the model for $g(a\mid w)$.
\end{theorem}
An immediate consequence of Theorem \ref{theo:eif} is that, in a randomized
trial, $S^{j,A}_{\P,\delta}(o) = 0$ for $j = 1,2$; however, even in such trials,
covariate adjustment can improve the efficiency of the resultant
estimator~\citep{vdl2003unified}. We now present the efficient scores
$S^{j,A}_{\P,\delta}(o)$ for modified treatment policies and exponentially
tilted stochastic interventions. To do so, we define the parameter $\q(a,w)
= \int \uu(z, a, w)\dd\P(z\mid w)$.
\begin{lemma}[Modified treatment policies]\label{lemma:mtp}
  If the modified treatment policy $d(A,W)$ satisfies Assumption~\ref{ass:inv},
  then
  \begin{align}
    S^{1,A}_{\P, \delta}(o) &= \bar \uu(d(a,w), w) - \int \bar
                                \uu(d(a,w), w)\g(a\mid w)\dd\kappa(a)\\
    S^{2,A}_{\P, \delta}(o) &= \q(d(a,w), w) - \int \q(d(a,w), w)
                                \g(a \mid w) \dd\kappa(a).
  \end{align}
\end{lemma}

\begin{lemma}[Exponential tilt]\label{lemma:tilt}
  If the stochastic intervention is the exponential tilt (\ref{eq:tilt}), then
  \begin{align}
    S^{1,A}_{\P, \delta}(o) &= \frac{\g_\delta(a \mid
                                w)}{\g(a \mid w)}\left\{\bar\uu(a, w) - \int
                                \bar\uu(a, w)\g_\delta(a\mid
                                w)\dd\kappa(a)\right\}\label{eq:scorea1}\\
    S^{2,A}_{\P, \delta}(o) &= \frac{\g_\delta(a \mid
                                w)}{\g(a \mid w)}\left\{\q(a, w) - \int
                                \q(a, w)\g_\delta(a\mid w)
                                \dd\kappa(a)\right\}\label{eq:scorea2}
  \end{align}
\end{lemma}

For binary exposures, the EIF for the incremental propensity score intervention
may be simplified as follows.
\begin{lemma}[EIF for incremental propensity score interventions]\label{coro:tilt}
  Let $A \in \{0, 1\}$ and let the exponentially tilted intervention
  $g_{\delta,0}(1\mid W)$ be based on (\ref{eq:tilt}) under the
  parameterization $\delta' = \exp(\delta)$. Then, the EIF of
  Lemma~\ref{lemma:tilt} may be simplified as follows. Specifically, we have
  $$S^{j,A}_{\eta,\delta}(o)= \frac{\delta\q^j(w)\{a -
  \g(1\mid w)\}}{\{\delta \g(1\mid w) + 1 - \g(1\mid w)\}^2}, \quad
  \text{where}$$
\begin{equation}
    \q^1(w) = \bar\uu(1, w) - \bar\uu(0, w), \quad \text{and} \quad
    \q^2(w) = \E \left\{\uu(Z,1, W) - \uu(Z,0, W) \mid W = w \right\}.
  \label{eq:defqs}
\end{equation}
\end{lemma}

In contrast to the efficient influence function for the interventional
(in)direct effects~\citep{diaz2020nonparametric}, the contribution of the
exposure mechanism to the EIF for the stochastic interventional effects is
nonzero. This is a direct consequence of the fact that the parameter of
interest depends on $\g(a \mid w)$; moreover, this implies that the efficiency
bound in observational studies differs from the efficiency bound in randomized
trials. Thus, it is not generally possible to obtain estimating equations robust
to inconsistent estimation of $\g(a \mid w)$. Such robustness will only be
possible if the stochastic intervention is also a modified treatment policy
satisfying Assumption~\ref{ass:inv}.

The form of Theorem~\ref{theo:eif} makes it clear that estimation of
multivariate or continuous conditional density functions on the mediators $Z$ or
intermediate confounders $L$, as well as integrals with respect to these density
functions, is generally necessary for computation of the EIF. This poses
a significant challenge from the perspective of estimation, due to both the
curse of dimensionality and the practical computational complexity inherent in
solving multivariate numerical integrals. A simplification is possible when
either either of $Z$ or $L$ is low-dimensional; this is achieved by
re-parameterizing the densities as conditional expectations (or low-dimensional
conditional densities) that take other nuisance parameters as pseudo-outcomes.
In cases where $L$ or $Z$ is low-dimensional, our proposed re-parameterizations
allow for the conditional density to be estimated via appropriate semiparametric
density estimation procedures~\citep[e.g.,][]{diaz2011super,
hejazi2021haldensify}.

\begin{lemma}[EIF for low-dimensional $L$]\label{lemma:altres}
  Let $L$ be low-dimensional and $Z$ multivariate. A representation of $\vv$,
  $\s$, and $\bar\uu$ in terms of conditional expectations may be chosen in
  order to simplify their estimation. Denote by $\b(l\mid a,w)$ and $\d(l\mid
  z,a,w)$ the density of $L$ conditional on $(A,W)$ and $(Z,A,W)$, respectively.
  Then, using (\ref{eq:parameterize}), we have
  \begin{align}
    \vv(l,a,w) & = \E\left[\m(z,l,a,w)\frac{\b(L\mid A, W)}{\d(L\mid Z, A,
                 W)}\,\bigg|\,L=l, A=a, W=w\right],\notag\\
    \s(l,a,w) & =\E\left[\m(z,l,a,w)\frac{\b(L\mid A, W)}{\d(L\mid Z, A,
                W)}\frac{\g(A\mid W)}{\e(A\mid Z, W)}\,\bigg|\,L=l, A=a,
                W=w\right],\label{eq:altnuis}\\
    \bar \uu(a,w) & = \E\left[\uu(Z,A,W)\,\bigg|\,A=a, W=w\right].\notag
  \end{align}
  Likewise,
  \begin{equation*}
    H^1_{\P,\delta}(a,z,l,w)  =  \frac{\g_\delta(a\mid
      w)}{\g(a\mid w)}\frac{\b(l\mid a, w)}{\d(l\mid z, a,
      w)}, \quad \text{and} \quad H^2_{\P,\delta}(a,z,l,w) =
      \frac{\g_\delta(a\mid w)}{\e(a\mid z,w)}\frac{\b(l\mid a, w)}
      {\d(l\mid z, a, w)}, \quad \text{then}
  \end{equation*}
  $$\q(a, w) = \E\left\{\frac{\g(A \mid W)}{\e(A \mid Z, W)} \uu(Z, A, W)
  \bigg| A = a, W = w \right\}.$$ Analogous representations may be constructed
  for $\bar \vv$, $\bar\s$, and $\uu$ based on the parameterizations
  (\ref{eq:parameterize}) if $L$ is multivariate and $Z$ is of low dimension. We
  note, however, that at least one of $Z$ or $L$ must be of low dimensionality
  for its density to be easily estimable and integrals over its range computed
  with relative ease. Henceforth, denote by $\eta = (\m, \g, \b, \bar\uu, \vv,
  \d, \e, \s, \q)$ and let $D_{\P,\delta}^j(o) = D_{\eta,\delta}^j(o)$.
\end{lemma}

We note that the choice of parameterization in Lemma~\ref{lemma:altres} has
important consequences for the purpose of estimation, as it helps to bypass
estimation of the (possibly high-dimensional) conditional density of the
mediators, by requiring only that the intermediate confounders be of modest
dimensionality for the purpose of estimation. In the particularly simple case
that $L=l\in\{0,1\}$ (as in our motivating application), the nuisance quantities
$\b(l\mid a,w)$ and $\d(l\mid z,a,w)$ reduce to conditional expectations,
allowing for regression methods, far more readily available throughout the
statistics literature and software, to be used for their estimation.
In addition to the expression for the EIF in Lemma~\ref{lemma:altres}, it is
important to understand the behavior of the difference $\P D_{\eta_1} - \theta$,
which is expected to yield a second-order term in differences $\eta_1-\eta$, so
that consistent estimation of $\theta$ is possible under consistent estimation
of certain configurations of the parameters in $\eta$. As we will see in
Theorems~\ref{theo:asos} and~\ref{theo:astmle}, this second-order term is
fundamental in the construction of asymptotically linear estimators.
Lemmas~\ref{supp:lemma:so1} and~\ref{supp:lemma:so2}, in the
\href{sm}{Supplementary Materials}, delineate these second-order terms. The
following lemma is a consequence.

\begin{lemma}[Multiple robustness for modified treatment
  policies]\label{lemma:dr1}
  Let the modified treatment policy satisfy \ref{ass:inv}, and let $\eta_1$ be
  such that one of the following conditions hold:
  \begin{table}[H]
    \centering
    \begin{tabular}{|c|c|c|c|c|c|c|c|c|c|}\hline
              & $\m_1=\m$ & $\g_1=\g$ & $\b_1=\b$ & $\bar\uu_1=\bar\uu$ & $\vv_1=\vv$ & $\d_1=\d$ & $\e_1=\e$ & $\s_1=\s$ & $\q_1=\q$ \\\hline
      Cond.~1 & $\times$  & $\times$  & $\times$  &                     &             &           &           &           &           \\
      Cond.~2 & $\times$  & $\times$  &           &                     & $\times$    &           &           & $\times$  &           \\
      Cond.~3 &           & $\times$  & $\times$  &                     &             & $\times$  & $\times$  &           &           \\
      Cond.~4 &           & $\times$  &           & $\times$            & $\times$    & $\times$  & $\times$  &           &           \\
      Cond.~5 & $\times$  &           & $\times$  & $\times$            &             &           &           &           & $\times$  \\
      Cond.~6 & $\times$  &           &           & $\times$            & $\times$    &           &           & $\times$  & $\times$  \\\hline
    \end{tabular}
    \caption{Different configurations of consistency for nuisance
      parameters}
    \label{tab:dr1}
  \end{table}
  Then $\P D_{\eta_1,\delta}^1=\theta_{1,\delta}$ and $\P
  D_{\eta_1,\delta}^2=\theta_{2,\delta}$, with $D_{\eta,\delta}^1$ and
  $D_{\eta,\delta}^2$ as defined in Theorem~\ref{theo:eif} and
  Lemma~\ref{lemma:mtp}.
\end{lemma}
The above lemma implies that it is possible to construct consistent estimators
for the (in)direct effects under consistent estimation of subsets of the
nuisance parameters in $\eta$, in the configurations described in the lemma.
Lemma~\ref{lemma:dr1} follows directly from Lemma~\ref{supp:lemma:so1}, found in
the \href{sm}{Supplementary Materials}. It may be surprising that estimation of
$\theta_{j,\delta}$ can be robust to inconsistent estimation of $\g$, even when
the parameter definitions are explicitly dependent on $\g$. We offer some
intuition for this result by noting that Assumption~\ref{ass:inv} allows use of
the change of variable formula to obtain $\theta_{2,\delta} = \E\left\{\int
\m(z, l, d(A,W),W)\p(l\mid d(A,W), W)\p(z \mid W)\dd\nu(z,l)\right\}$.
Estimation of this parameter without relying on $\g$ may be carried out by
consistently estimating $\m(z,l,a,w)$, $\p(l\mid a, w)$, and $\p(z\mid w)$ and
using the empirical distribution as an estimator of the outer expectation. This
behavior has been previously observed for related modified treatment policy
effects~\ref{ass:inv}~\citep{diaz2012population, haneuse2013estimation,
diaz2020causal}.

Robustness for exponentially tilted interventions (\ref{eq:npsem}), not
satisfying Assumption~\ref{ass:inv}, appears in Lemma~\ref{lemma:dr2}.
\begin{lemma}[Multiple robustness for exponential tilting]
  \label{lemma:dr2}\label{finallemma}
  Let $\g_\delta$ be defined as in (\ref{eq:tilt}) and $\eta_1$ be such that
  one or more of Cond.~1-4 in Table~\ref{tab:dr1} holds, then $\P
  D_{\eta_1,\delta}^1=\theta_{1,\delta}$ and $\P
  D_{\eta_1,\delta}^2=\theta_{2,\delta}$, with $D_{\eta,\delta}^1$ and
  $D_{\eta,\delta}^2$ as defined in Theorem~\ref{theo:eif} and
  Lemma~\ref{lemma:tilt}
\end{lemma}
Lemma~\ref{lemma:dr2} is a direct consequence of Lemma~\ref{supp:lemma:so2} in
the \href{sm}{Supplementary Materials}. The corresponding proof reveals that the
EIF for the binary distribution is not robust to inconsistent estimation of $\g$
--- that is, the intervention fails to satisfy Assumption~\ref{ass:inv} and
integrals over the range of $A$ cannot be computed using the change of variable
formula. This behavior has been previously observed for other interventions that
do not satisfy Assumption~\ref{ass:inv}. Even though this lemma implies that
consistent estimation of $\g$ is required, the bias terms remain second-order;
thus, an estimator of $\g$ converging at rate $n^{1/4}$ or faster is sufficient.

\section{Efficient estimation and statistical inference}\label{sec:estima}
We discuss two efficient estimators that rely on the efficient influence
function $D_{\eta, \delta}$, in order to build an estimator that is both
asymptotically efficient and robust to model misspecification. We discuss an
asymptotic linearity result for the doubly robust estimator that allows
computation of asymptotically accurate Wald-style confidence intervals and
hypothesis tests. In the sequel, we assume that preliminary estimators of the
components of $\eta$ are available. These estimators may be obtained from
flexible regression techniques such as neural networks, regression trees,
boosting, splines, or ensembles thereof~\citep{breiman1996stacked,
vdl2007super}. The consistency of these estimators determines consistency of
our estimators of $\theta_{j, \delta}$.

Both of our proposed efficient estimators make use of the EIF $D_{\eta, \delta}$
to revise an initial substitution estimator through a bias correction step.
Estimation proceeds by first constructing initial estimators of the
nuisance parameters in $\eta$; then, each of the efficient estimators is
constructed by application of distinct bias-correction steps. In this process,
we advocate for the use of cross-fitting~\citep{klaassen1987consistent,
zheng2011cross} to avoid imposing entropy conditions on
the initial estimators of the nuisance parameters in $\eta$. Let ${\cal V}_1,
\ldots, {\cal V}_J$ denote a random partition of the index set $\{1, \ldots,
n\}$ into $J$ prediction sets of approximately the same size. That is, ${\cal
V}_j\subset \{1, \ldots, n\}$; $\bigcup_{j=1}^J {\cal V}_j = \{1, \ldots, n\}$;
and ${\cal V}_j\cap {\cal V}_{j'} = \emptyset$. For each $j$, the associated
training sample is given by ${\cal T}_j = \{1, \ldots, n\} \setminus {\cal
V}_j$, and we let $j(i)$ denote the index of the validation set containing
observation $i$. Denote by $\hat \eta_{j}$ the estimator of $\eta$ obtained by
training a prediction algorithm using only data in the sample ${\cal T}_j$.

\subsection{Efficient One-Step Estimator}

To construct a robust and efficient estimator using the efficient influence
function $D_{\eta, \delta}$, the one-step bias
correction~\citep{pfanzagl1985contributions, bickel1993efficient} adds the
empirical mean of the estimated EIF $D_{\hat{\eta}, \delta}$ to an initial
substitution estimator. The estimators are thus defined
\begin{equation}\label{eq:aipw}
    \psidos = \frac{1}{n} \sum_{i = 1}^n \{D^1_{\hat\eta_{j(i)},
      0}(O_i) - D^2_{\hat\eta_{j(i)}, \delta}(O_i)\} \quad \text{and} \quad
    \psiios = \frac{1}{n} \sum_{i = 1}^n \{D^2_{\hat\eta_{j(i)},
      \delta}(O_i) - D^1_{\hat\eta_{j(i)}, \delta}(O_i)\}.
\end{equation}
Asymptotic linearity and efficiency of estimators for modified treatment
policies follows.

\begin{theorem}[Weak convergence of one-step estimators]\label{theo:asos}
  Let $\norm{\cdot}$ denote the $L_2(\P)$-norm defined as
  $\norm{f}^2 = \int f^2 \dd \P$. Define the following conditions.
  \begin{enumerate}[label=(C\arabic*)]
  \item \label{ass:bounded}
    $\P\{|D_{\eta, \delta}^j(O)| \leq C \} = \P \{| D_{\hat{\eta},
      \delta}^j(O) | \leq C \} = 1$ for $j=1,2$ and for some
    $C < \infty$.
  \item The following second-order terms converge at the specified rate
    \label{ass:sec1}
    $\norm{\hat{\m} - \m} \{\norm{\hat{\g} - \g} +
        \norm{\hat{\e} - \e} + \norm{\hat{\d} - \d}\} = o_\P(n^{-1/2})$,
  $\norm{\hat{\g} - \g}\{\norm{\hat{\bar\uu} -
      \bar\uu} + \norm{\hat{\q} - \q}\} = o_\P(n^{-1/2})$,
  $\norm{\hat{\b} - \b}\{\norm{\hat{\vv} - \vv} +
    \norm{\hat{\s} - \s}\} = o_\P(n^{-1/2})$.
  \item \label{ass:pwinv} The effect is defined in terms of modified treatment
    policy $d(a,w)$, which is piecewise smooth invertible (\ref{ass:inv}).
  \item \label{ass:gconv} The intervention $\g_\delta$ is an
    exponential tilting intervention and
    $\P\left\{ \int(\hat{\g} - \g) \dd\kappa \right\}^2 =
    o_\P(n^{-1/2})$.
  \end{enumerate}
  If Conditions~\ref{ass:bounded} and~\ref{ass:sec1} hold, and one of
  Conditions~\ref{ass:pwinv} and~\ref{ass:gconv} holds, then:
    $\sqrt{n}\{\psidos - \psid\} \rightsquigarrow N(0,
    \sigma^2_{D,\delta})$, and
    $\sqrt{n}\{\psiios - \psii\} \rightsquigarrow N(0,
                                  \sigma^2_{I,\delta})$,
  where
  $\sigma^2_{D,\delta} = \var\{D_{\eta, 0}^1(O)
  -D_{\eta, \delta}^2(O)\}$ and $\sigma^2_{I,\delta} =
  \var\{D_{\eta, \delta}^2(O) - D_{\eta, \delta}^1(O)\}$ are the respective
  efficiency bounds.
\end{theorem}

Theorem~\ref{theo:asos} establishes the weak convergence of $\psidos$ and
$\psiios$ pointwise in $\delta$. This convergence is useful to derive confidence
intervals in situations where the MTP has a scientific interpretation for
a given realization of $\delta$. Under Theorem~\ref{theo:asos}, an estimator
$\hat\sigma^2_{D,\delta}$ of $\sigma^2_{D,\delta}$ may be obtained as the
empirical variance of $D_{\hat\eta_{j(i)}, 0}^1(O_i) - D_{\hat\eta_{j(i)},
\delta}^2(O_i)$, and a Wald-style confidence interval may be constructed as
$\psidos\pm z_{1-\alpha/2} \hat\sigma^2_{D}(\delta)/\sqrt{n}$; the same applies
to $\psiios$.

Although the one-step estimator has optimal asymptotic performance, its
finite-sample behavior may be affected by the inverse probability weighting
involved in the computation of the efficient influence functions
$D_{\hat\eta}^j(O_i):j=1,2$. In particular, it is not guaranteed that $\psidos$
and $\psiios$ will remain within the bounds of the parameter space. This issue
may be attenuated by performing weight stabilization. The estimated EIF
$D_{\hat\eta_{j(i)}}^1(O_i)$ can be weight-stabilized by dividing
(\ref{eq:scorey1}) and (\ref{eq:scorey2}) by the empirical mean of
$H_{\hat\eta_j(i),\delta}^1(A_i,Z_i,L_i,W_i)$ and
$H_{\hat\eta_j(i),\delta}^2(A_i,Z_i,L_i,W_i)$, respectively; as well as dividing
(\ref{eq:scorelz1}), (\ref{eq:scorelz2}), (\ref{eq:scorea1}), and
(\ref{eq:scorea2}) by the empirical mean of $\hat \g_{j(i), \delta}(A_i\mid
W_i)/\hat \g_{j(i)}(A_i\mid W_i)$.

\subsection{Efficient Targeted Minimum Loss Estimator}

Although corrections may be applied to the one-step estimator, a more principled
way to obtain estimators that remain in the parameter space may be derived from
the targeted minimum loss (TML) estimation framework. The TML estimator is
constructed by tilting an initial data adaptive estimator $\hat{\eta}$ towards
a solution $\tilde{\eta}$ of the estimating equations
\begin{equation}
    \Pn \{D_{\tilde\eta,0}^1 - D_{\tilde\eta,\delta}^2\} =
    \psid(\tilde\eta) \quad \text{and} \quad
    \Pn \{D_{\tilde\eta,\delta}^2 - D_{\tilde\eta,\delta}^1\} =
    \psii(\tilde\eta),
  \label{eq:eqtmle}
\end{equation}
where $\psid(\tilde\eta)$ and $\psii(\tilde\eta)$ are the substitution
estimators in formula~\eqref{eq:tmle} obtained by plugging in the estimates
$\tilde{\eta}$ in the parameter definition~\eqref{eq:defpsi}. Thus, a TML
estimator is guaranteed to remain in the parameter space by virtue of its being
a substitution estimator. The fact that the nuisance estimators solve the
relevant estimating equation is used to obtain a weak convergence result
analogous to Theorem~\ref{theo:asos}. Thus, while the TML estimator is expected
to attain the same optimal asymptotic behavior as the one-step estimator, its
finite-sample behavior may be better. An algorithm to compute a TML estimator
$\tilde\eta$ is presented in the \href{sm}{Supplementary Materials}. Roughly,
the algorithm proceeds by projecting the EIF into score functions for the model
of each nuisance parameter and fitting appropriate parametric
submodels~\citep{vdl2011targeted}. For example, the following
model is fitted for $\m$:
$$\logit \m_\beta(a,z,l,w) = \logit \hat
  \m(z,l,a,w) + \beta_I H_{I}(o) +
  \beta_D H_{D}(o), \quad \text{where}$$
\begin{equation*}
  H_{ D}(o) = \frac{{\hat\b}(l\mid a,
                                w)}{{\hat\d}(l \mid z, a,
                                w)}\left\{1-\frac{{\hat\g}_\delta(a\mid
                                w)}{{\hat\e}(a\mid z,w)}\right\},
  \quad \text{and} \quad
  H_{I}(o) = \frac{{\hat\b}(l\mid a,
                                w)}{{\hat\d}(l \mid z, a,
                                w)}\left\{\frac{{\hat\g}_\delta(a\mid
                                w)}{{\hat\e}(a\mid
                                z,w)}-\frac{\hat{\g}_\delta(a
                                \mid w)}{{\hat\g}(a\mid w)}\right\},
\end{equation*}
and $\logit(p) = \log\{p(1-p)^{-1}\}$. Here, the initial estimator $\logit
\hat\m(z,l,a,w)$ is considered a fixed offset variable (i.e., a variable with
known parameter value equal to one). The score of these tilting models is equal
to the corresponding component of the EIF. The parameter $\beta=(\beta_I,
\beta_D)$ may be estimated via standard logistic regression of $Y$ on
$(H_{D}(O), H_{I}(O))$ with no intercept and an offset term equal to $\logit
\hat\m(z,l,a,w)$. Let $\hat\beta$ denote the MLE, and let $\tilde
\m=\m_{\hat\beta}$ denote the updated estimates. Fitting this regression model
ensures that $\tilde\m$ solves the relevant score equations. Regression models
like this are estimated iteratively for all parameters in a way that guarantees
that the estimating equations (Equation~\eqref{eq:eqtmle}) are solved up to an
error term that converges to zero in probability at rate faster than $n^{-1/2}$.
Upon termination of the iterative process, the TML estimators are defined as
\begin{equation}
  \begin{split}
    \psidtmle &= \frac{1}{n} \int\sum_{i = 1}^n
    \left\{\tilde{\bar\uu}(a,W_i)\tilde\g(a\mid W_i) -
      \tilde\uu(Z_i,a,W_i)\tilde\g_\delta(a\mid W_i)\right\}\dd\kappa(a)\\
    \psiitmle &= \frac{1}{n} \int\sum_{i = 1}^n
    \left\{\tilde\uu(Z_i,a,W_i) -
      \tilde{\bar\uu}(a,W_i)\right\}\tilde\g_\delta(a\mid
    W_i)\dd\kappa(a).
  \end{split}\label{eq:tmle}
\end{equation}
The fact that the TML estimator solves estimating equations
(Equation~\eqref{eq:eqtmle}) is fundamental to the following theorem.
\begin{theorem}[Weak convergence of TML estimator]
  \label{theo:astmle}\label{finalthm}
  Assuming that~\ref{ass:bounded} and~\ref{ass:sec1} hold, and one
  of~\ref{ass:pwinv} and~\ref{ass:gconv} of Theorem~\ref{theo:asos} hold, then
    $\sqrt{n}\{\psidtmle - \psid\} \rightsquigarrow N(0,
                                  \sigma^2_{D,\delta}),
    \sqrt{n}\{\psiitmle - \psii\} \rightsquigarrow N(0,
                                  \sigma^2_{I,\delta})$,
  where
  $\sigma^2_{D,\delta} = \var\{D_{\eta, 0}^1(O)
    - D_{\eta, \delta}^2(O)\}$ and $\sigma^2_{I,\delta} =
    \var\{D_{\eta, \delta}^2(O) - D_{\eta, \delta}^1(O)\}$.
\end{theorem}
Using Theorem~\ref{theo:astmle}, asymptotically valid variance estimators,
p-values, and confidence intervals for the (in)direct effects may be obtained in
a manner analogous to those for the one-step estimator. The proof of the theorem
proceeds using similar arguments as the proof of Theorem~\ref{theo:asos} for the
one-step estimator, using empirical process theory and leveraging cross-fitting
to avoid entropy conditions on the initial estimators of $\eta$. Since the
estimators now depend on the full sample through the estimates of the parameters
$\beta$ of the logistic tilting models, the empirical process treatment differs
slightly to that of Theorem~\ref{theo:asos}; its proof is detailed in the
\href{sm}{Supplementary Materials}.

In Section~\ref{supp:sec:sim} of the \href{sm}{Supplementary Materials}, we
present a simulation study comparing the two efficient estimators under
different configurations of nuisance parameter misspecification. In brief, our
findings illustrate that our estimators empirically satisfy the forms of
robustness identified by our theoretical investigations; moreover, in keeping
with prior investigations in other settings~\citep[e.g.,][]{vdl2011targeted},
the TML estimator generally outperforms the one-step estimator in terms of both
bias and efficiency. Given the favorable evaluation of our proposed estimators
in these experiments, we next demonstrate their application.

\section{Application to the X:BOT trial}\label{sec:applic}

We now apply our stochastic interventional direct and indirect effects to
decompose the causal effect of a strategy where buprenorphine dose is
successively increased early in the treatment course (regardless of opioid use)
on relapse among those with opioid use disorder (OUD). Data for our illustrative
analysis come from the X:BOT trial, a 24-week, multi-site randomized controlled
trial designed to examine the comparative effectiveness of extended-release
naltrexone (XR-NTX) and sublingual buprenorphine-naloxone (BUP-NX) on
relapse~\citep{lee2018comparative}. The X:BOT trial enrolled 570 participants,
all of whom were 18 years or older, had OUD~\citep[as per the Diagnostic and
Statistical Manual of Mental Disorders-5;][]{apa2013dsm5}, and had used
non-prescribed opioids in the 30 days preceding enrollment. Participants were
randomized to receive either XR-NTX or BUP-NX using a stratified permuted block
design; 287 of the 570 were randomized to receive BUP-NX. Prior analytic efforts
have established a protective effect of BUP-NX administration (versus placebo)
on OUD relapse~\citep{mattick2014buprenorphine}. For each participant assigned
to receive BUP-NX, the prescribed dose was based on both clinical
indication~\citep{lee2018comparative} and clinician judgment. Some clinicians
tended to hold dose constant over time (i.e., a static regimen), while others
increased dose --- either based on clinical assessment or on the hypothesis that
higher doses would result in better outcomes~\citep{comer2005buprenorphine}. We
estimated stochastic interventional (in)direct effects to assess the mechanism
by which universally ramping up BUP-NX dose early in treatment (i.e., three or
more dose increases in the first four weeks of treatment) could mitigate the
risk of OUD relapse.

\begin{figure}[h!]
  \centering
  \includegraphics[scale=0.3]{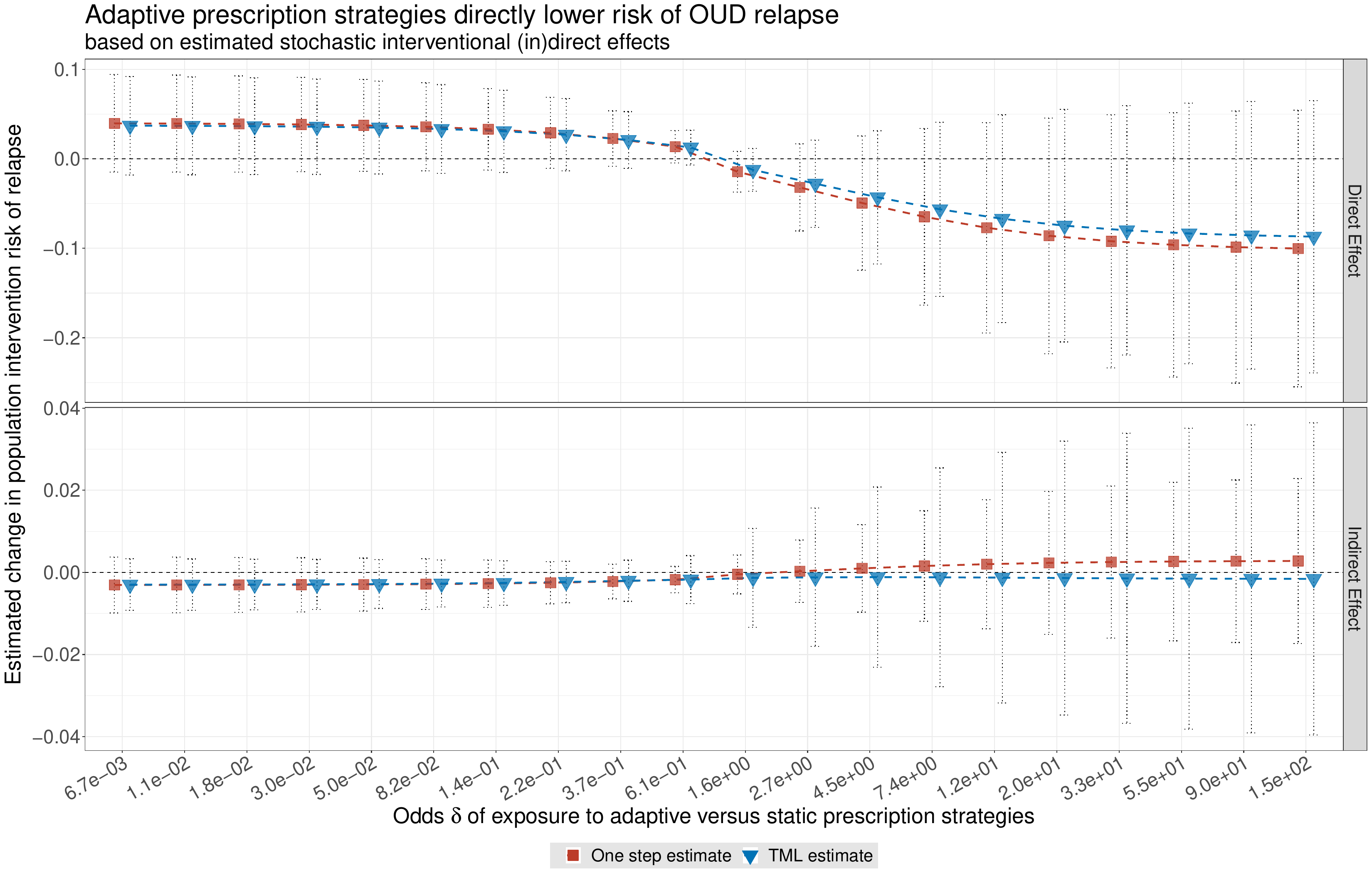}
  \caption{Stochastic interventional direct (upper panel) and indirect (lower
  panel) effect estimates of hypothetically increasing exposure odds to a
  BUP-NX dose schedule in which dose is increased early in OUD relapse
  treatment.}
  \label{fig:xbot_ipsi}
\end{figure}

Baseline covariates ($W$) available in the data included site; gender; age;
race/ethnicity; homeless status; educational attainment; employment status;
marital status; current intravenous drug use; alcohol use disorder; cocaine use
disorder; age at start of heroin use; severity of current opioid use; indicator
of prior OUD treatment; past withdrawal discomfort level; histories of
amphetamine use, sedative use, and cannabis use; weekly cost of primary drug;
whether or not living with an individual currently using drugs or with alcohol
use disorder; histories of psychiatric illnesses; randomization timing; baseline
pain level; baseline depression symptoms. The exposure ($A$) was taken to be
successive increases in dose of BUP-NX versus static dose, measured during the
first four weeks of treatment. Mediating factors ($Z$) included depression and
pain, measured from week 6 until relapse or week 24 (end of follow-up).
Abstinence from illicit opioid use early in the treatment schedule, measured
between weeks 4 and 6, acted as an intermediate confounder affected by exposure
($L$). OUD relapse status at the X:BOT trial's end of follow-up was the outcome
of interest ($Y$). To examine the effect of exposure to successive increases in
BUP-NX dose, we consider an incremental propensity score intervention, which,
for binary $A$, replaces the propensity score $g(1 \mid w)$ with a shifted
variant constructed from multiplying the odds of exposure by a user-specified
degree parameter $\delta$~\citep{kennedy2019nonparametric}, which we vary along
a grid $\log(\delta) \in \{-10.0, -9.5, \ldots, 9.5, 10.0\}$ of the observed
exposure odds.
Across all such estimates in the odds $\delta$ of exposure, the stochastic
interventional (in)direct effects that we estimated may be interpreted
in terms of the overall effect of increasingly encouraging ramping up BUP-NX
dose early in treatment on the counterfactual risk of OUD relapse; thus, the
results of our analysis may be informative of the mechanisms by which increasing
BUP-NX dose can alter the risk of OUD relapse. Figure~\ref{fig:xbot_ipsi}
presents the direct and indirect effect estimates across the grid in $\delta$.
We applied both of our cross-fitted, efficient one-step and TML estimators to
examine the stochastic interventional direct and indirect effects of increasing
the odds of ramping up BUP-NX dose. Both strategies produced results generally
in close agreement with the magnitude of the (in)direct effects. For each point
estimate, standard error estimates and 95\% Wald-style confidence intervals were
constructed based on Theorem~\ref{theo:astmle}. To ensure flexibility of our
estimators, all nuisance parameters $\eta = (\e, \m, \d, \g, \b, \uu, \vv, \s,
\q)$ were estimated using the Super Learner ensemble modeling
algorithm~\citep{vdl2007super,coyle2021sl3}.
From examination of the point estimates and confidence intervals of the
(in)direct effects in Figure~\ref{fig:xbot_ipsi}, two conclusions may be drawn.
Firstly, there appears to be little to no indirect effect of successively
increasing BUP-NX dose on risk of OUD relapse, revealing that any effect of
BUP-NX dose cannot be explained by actions on mediating factors such as pain or
depression. Secondly, the direct effect of successively increasing BUP-NX dose
varies considerably across changes in the odds of the introduction of such
a dosing schedule. Importantly, lowered odds of dosage increases could lead to
as much as a 5\% increase in OUD relapse risk, with a plateau emerging at odds
lower than $\approx$0.1\%, suggesting that static dose can lead to comparatively
heightened relapse risk. OUD relapse risk further appears to decrease by
$\approx$10\% with increased odds of successive BUP-NX dose increases, with the
risk plateauing at odds higher than 33\%. This decrease in the counterfactual
risk of OUD relapse suggests a protective effect of BUP-NX dose schedules when
dose is successively increased early in the treatment course; however, the
action mechanism is not readily apparent.

The conclusions that may be drawn from our re-analysis using the stochastic
interventional (in)direct effects complement those previously reported in the
investigations of~\citet{lee2018comparative}, who evaluated the total effect of
BUP-NX (versus XR-NTX) treatment on OUD relapse,
and~\citet{rudolph2020explaining}, who used the interventional mediation
analysis approach of~\citet{diaz2020nonparametric} (for static interventions on
$A$) to examine differences in relapse risk between homeless and non-homeless
participants. Importantly, our substantive conclusion --- that dosage increases
directly lower the risk of relapse --- agrees with those
of~\citet{rudolph2020association}, who found that dosage increases directly
lowered risk of OUD relapse when such increases followed opioid use. Notably,
our proposed (in)direct effects and estimation approach differ from prior
efforts in three important ways: (i) our causal effect definitions remain
unaltered in the presence of intermediate confounders affected by exposure and
may be re-evaluated in randomized trials, (ii) the flexible estimators we
introduce eschew restrictive modeling assumptions by incorporating modern
machine learning in nuisance parameter estimation, and (iii) our strategy
provides an analog to a dose-response analysis by allowing for the risk of OUD
relapse to be traced out across changes in the odds of exposure to a schedule in
which BUP-NX dose is increased repeatedly early in treatment.

\section{Discussion}\label{sec:discuss}


We have proposed a class of novel direct and indirect effect estimands for
causal mediation analysis, as well as two efficient estimators of these effects
in the nonparametric statistical model. Importantly, our proposed estimation
framework allows for data adaptive estimation of nuisance parameters, while
still preserving the benefits associated with similar classical techniques: our
estimators are regular and asymptotically linear, provide unbiased point
estimates, are multiply robust, allow the construction of asymptotically valid
confidence intervals, and are capable of attaining the nonparametric efficiency
bound. Notably, our (in)direct effects remain well-defined even in the presence
of intermediate confounders affected by exposure. Further, any scientific
conclusions drawn based upon our proposed (in)direct effects may be readily
interrogated in trials that randomize both the exposure and mediators. Such
flexible effect definitions and estimators appear necessary both to cope with
the design complexity of modern epidemiological and biomedical studies and to
take advantage of the ever-growing number of flexible, data adaptive regression
techniques.

The challenge of leveraging data adaptive regression methodology to construct
robust estimators that accommodate valid statistical inference is not a new one.
It has been considered in great detail as early as the work
of~\citet{pfanzagl1985contributions} as well in numerous recent advances, most
notably by~\citet{vdl2011targeted, vdl2018targeted}
and~\citet{chernozhukov2018double}; related work by these authors presents
a wealth of extensions and applications. In the present work, we derive multiply
robust, efficient estimators based on both the one-step and targeted minimum
loss estimation frameworks. Following~\citet{klaassen1987consistent}
and~\citet{zheng2011cross}, our estimators leverage cross-validation to avoid
imposing possibly restrictive assumptions on nuisance function estimators. We
demonstrated the properties of our estimators in simulation experiments that
illustrated their ability to yield unbiased point estimates, attain the
nonparametric efficiency bound, and build confidence intervals exhiting coverage
at the nominal rate across several nuisance parameter configurations --- all
within a context in which classical mediation effects are ill-defined. We
demonstrated the application of our novel (in)direct effects in dissecting the
mechanism by which increasing the odds of adopting a dosing schedule of
universal successive increases in buprenorphine early in treatment affects OUD
relapse~\citep{lee2018comparative, rudolph2020explaining}.

Several significant extensions and refinements are left for future
consideration. Firstly, our proposed estimation strategy for the direct and
indirect effects leverages re-parameterizations of factors of the likelihood in
order to simplify the estimation of nuisance parameters. This approach works
particularly well when either mediators or intermediate confounders are of
modest dimension; however, improvements can be made to accommodate settings in
which both mediators and intermediate confounders are of high dimensionality.
When defining effects based upon stochastic interventions indexed by the
user-specified parameter $\delta$, an important consideration is choosing
\textit{a priori} a particular value of $\delta$. One solution is to evaluate
a set of causal effects indexed by a grid in $\delta$. In such cases, aggregate
effects (across $\delta$) may be summarized via working marginal structural
models~\citep[e.g.,][]{hejazi2020efficient} or the construction of uniform tests
of the null hypothesis of no direct effect~\citep[e.g.,][]{diaz2020causal}.
Developments of these distinct summarization strategies would enrich the range
of scientific problems to which these robust and flexible direct and indirect
effects may be applied.

\section*{Supplementary Materials}

The reader is referred to the on-line Supplementary Materials for technical
appendices. \texttt{R} scripts used to conduct the simulation experiments and
real-world data analysis have been made publicly available in a GitHub
repository at \url{https://github.com/nhejazi/pub_medshift_interv_biostats}.
While the estimation machinery is accessible from that GitHub repository, its
integration into our open source \texttt{medshift} \texttt{R}
package~\citep{hejazi2020medshift} (\url{https://github.com/nhejazi/medshift})
is ongoing and will support wider long-term usage.

\section*{Acknowledgments}

The authors thank John Rotrosen, Edward Nunes, and Marc Fishman for raising the
research question in the Application and for helpful feedback. KER's time was
supported by a grant from the National Institute on Drug Abuse (award
no.~R00-DA042127), MJvdL's time was supported by a grant from the National
Institute of Allergy and Infectious Diseases (award no.~R01-AI074345), and NSH's
time was supported by a grant from the National Science Foundation (award
no.~DMS-2102840). The X:BOT trial was supported by the National Institute on
Drug Abuse, Clinical Trials Network (award no.'s U10DA013046, UG1/U10DA013035,
UG1/U10DA013034, U10DA013045, UG1/U10DA013720, UG1/U10DA013732, UG1/U10DA013714,
UG1/U10DA015831, U10DA015833, HHSN271201200017C, and HHSN271201500065C).

\bibliographystyle{plainnat}
\bibliography{refs}
\end{document}


\maketitle
\section{Simulation study}\label{sec:sim}

We used simulation experiments to assess our two proposed efficient estimators
of the (in)direct effects. On account of computational considerations, we focus
on binary exposures and intermediate confounders in this example; however, as
noted in the prior, our proposed methodology is general enough to be readily
applicable in the presence of continuous-valued covariates, treatment,
mediators, intermediate confounders, and outcome. We used the following
data-generating mechanism for the joint distribution of $O$ to generate
synthetic data to evaluate our two estimators:
\begin{align*}
W_1 &\sim \text{Bernoulli}(p = 0.6);\\
W_2 &\sim \text{Bernoulli}(p = 0.3);\\
W_3 \mid (W_1, W_2) &\sim \text{Bernoulli}(p = 0.2 + 1/3 \cdot (W_1 + W_2));\\
A \mid W &\sim \text{Bernoulli}(p = \expit(2 + (5 / (W_1 +
    W_2 + W_3))));\\
L \mid (A, W) &\sim \text{Bernoulli}(p = \expit(1/3(W_1 + W_2 + W_3) - A -
  \log(2) + 0.2));\\
Z \mid (L, A, W) &\sim \text{Bernoulli}(p = \expit(\log(3) \cdot (W_1 + W_2)
    + A - L));\\
Y \mid (Z, L, A, W) &\sim \text{Bernoulli}\left(p = \expit\left(1 -
   \frac{3 \cdot (3 - L - 3A + Z)}{2 + (W_1 + W_2 + W_3)} \right)\right),
\end{align*}
where $\expit(x) \coloneqq \{1 + \exp(x)\}^{-1}$. For each of the sample sizes
$n \in \{200, 800, 1800, 3200, 5000, 7200, \allowbreak 9800, 12800,
16200\}$, $500$ datasets were generated. For every dataset, six variations of
each of the two efficient estimators was applied --- five variants were based on
misspecification of a single nuisance parameter among $\{\e, \m, \d, \g, \b\}$
while the sixth variant was constructed based on consistent estimation of all
five nuisance parameters. An intercept-only logistic regression model
provided inconsistent estimation of each of the nuisance parameters $\{\e, \m,
\d, \g, \b\}$, while a Super Learner ensemble~\citep{vdl2007super} was used to
achieve consistent estimation. The Super Learner ensemble was constructed with
a library of algorithms composed of intercept-only logistic regression;
main-terms logistic regression; and several variants of the highly adaptive
lasso~\citep{benkeser2016highly, vdl2017generally, coyle2020hal9001},
a nonparametric regression approach capable of flexibly estimating arbitrary
functional forms at a fast convergence rate under only a global smoothness
assumption~\citep{vdl2017uniform, bibaut2019fast}.
Note that we do not consider cases of misspecified estimation of $\{\vv, \s,
\q, \bar\uu\}$, as their consistent estimation depends on a subset of the
nuisance parameters $\{\e, \m, \d, \g, \b\}$. Generally, based on
Lemmas~\ref{main:lemma:dr1} and~\ref{main:lemma:dr2}, robustness of the direct
and indrect effect estimators to misspecification of $\{\e, \m, \d\}$ is to be
expected, but the same is not true under misspecification of $\{\g, \b\}$.

\begin{figure}[H]
  \centering
  \includegraphics[scale = 0.45]{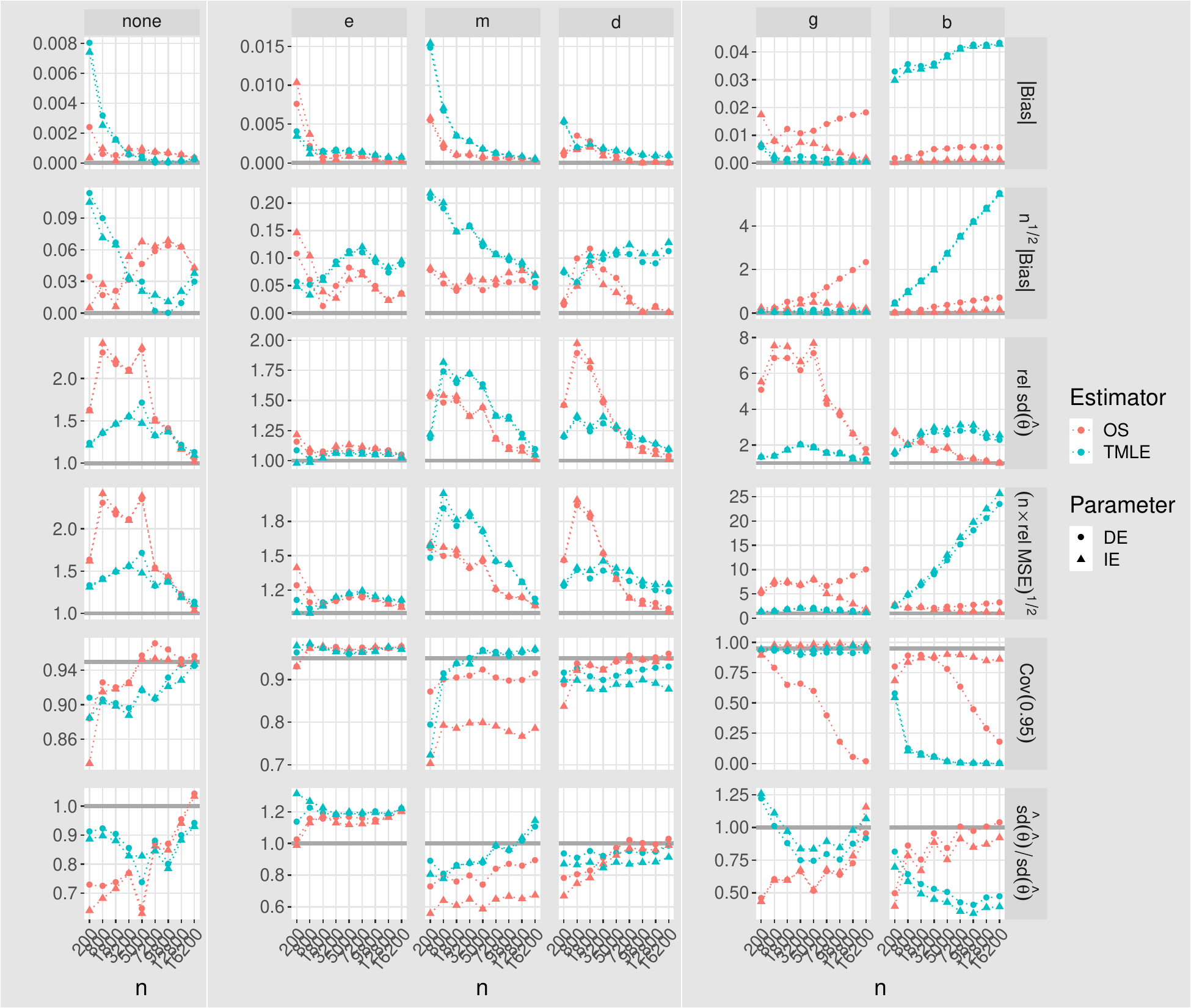}
  \caption{Comparison of efficient estimators across different nuisance
    parameter configurations.}\label{fig:simula}
\end{figure}

Figure~\ref{fig:simula} summarizes the results of our investigations of the
relative performance of the estimator variants enumerated above. Specifically,
we assess the relative performance of our proposed estimators in terms of
absolute bias, scaled (by $n^{1/2}$) bias, standard error and scaled (by $n$)
mean squared error relative to the efficiency bound for the data-generating
model, the empirical coverage of 95\% confidence intervals, and relative
efficiency. In terms of both raw (unscaled) bias and scaled bias, the estimator
variants appear to conform to the predictions of Lemmas~\ref{main:lemma:dr1}
and~\ref{main:lemma:dr2} --- specifically, raw bias vanishes and scaled bias
stabilizes to a small value (providing evidence for rate-consistency) under
misspecification of any of $\{\e, \m, \d\}$ as well as in the case of no
nuisance parameter misspecification. In the same vein, when either of $\{\g,
\b\}$ are estimated inconsistently, some of the estimator variants display
diverging asymptotic (scaled) bias, in agreement with expectations based upon
theory. The consistency of other estimator variants (e.g., the one-step
estimator under misspecification of $\b$) is likely an artifact of this
data-generating mechanism, not to be taken as a general indication of robust
performance. In terms of their relative mean squared error, the estimators of
the (in)direct effects exhibit convergence to the efficiency bound under
misspecification of $\{\e, \m, \d\}$ and under no misspecification; this also
appears to hold for a subset of the estimator variants under misspecification of
$\{\g, \b\}$. We stress that aspects of this are likely to be a particularity of
the given data-generating mechanism or on account of the irregularity of
misspecified estimator variants, for the regularity and asymptotically linearity
of the estimators is only to be expected under consistent estimation of all
nuisance parameters. Finally, the empirical coverage of 95\% confidence
intervals is as expected: under a lack of nuisance parameter misspecification,
both the one-step and TML estimators of the direct and indirect effect achieve
95\% coverage in larger sample sizes. We note that misspecification of $\e$
leads to over-coverage for all estimator variants, implying an overly inflated
variance estimate, while the confidence intervals fail to attain the nominal
rate in most other instances. Notably, several of the estimator variants
generate confidence intervals that are liable to converge to 0\% coverage in
larger samples under misspecification of $\{\g, \b\}$, very much in line with
theoretical expectations.

Importantly, the TML estimator appears to generally outperform the one-step
estimator throughout several scenarios. This comes in several forms, including
lower bias, relative standard deviation, or relative mean squared error under
misspecification of $\{\e, \m, \d\}$ or under no misspecification; however,
under inconsistent estimation of $\{\g, \b\}$, the irregularity of the
estimators complicates this comparison. Interestingly, under misspecification of
$\g$, the TML estimators of the direct and indirect effects appear unbiased and
efficient, a result unpredictable from theory given the irregularity of the
estimators under this configuration. Altogether, results of our numerical
experiments indicate that our proposed estimators exhibit properties that align
with the theoretical results of Lemmas~\ref{main:lemma:dr1}
and~\ref{main:lemma:dr2}.

In the interest of transparency and computational reproducibility, \texttt{R}
scripts used to conduct these simulation experiments and summarize their results
have been made publicly available in a curated GitHub repository located at
\url{https://github.com/nhejazi/pub_medshift_interv_biostats}. The estimation
machinery currently accessible via that repository is in the process of being
integrated into the open source \texttt{medshift} \texttt{R} package, available
at \url{https://github.com/nhejazi/medshift}~\citep{hejazi2020medshift}.

\section{Theorem \ref{main:theo:iden}}
\begin{proof}
  First, we have
  \begin{align}
    \E&\{Y_{A_\delta, G_\delta}\} \notag\\&= \int \E\left\{Y_{a,z}\mid
                                            A_\delta=a,G_\delta = z,
                                            W=w\right\}\g_\delta(a\mid
                                            w)\P(G_\delta=z\mid
                                            A_\delta=a,W=w)\p(w)\dd\nu(a,z,w)\notag\\
      &= \int \E\left\{Y_{a,z}\mid W=w\right\}\g_\delta(a\mid
        w)\P(Z(a)=z\mid
        A_\delta=a,W=w)\p(w)\dd\nu(a,z,w)\label{eq:t1e1}\\
      &= \int \E\left\{Y_{a,z}\mid A=a,W=w\right\}\g_\delta(a\mid
        w)\P(Z(a)=z\mid W=w)\p(w)\dd\nu(a,z,w)\label{eq:t1e2}\\
      &= \int \E\left\{Y_{a,z}\mid A=a,W=w\right\}\g_\delta(a\mid
        w)\P(Z(a)=z\mid A=a,W=w)\p(w)\dd\nu(a,z,w)\label{eq:t1e3}\\
      &= \int \E\left\{Y_{a,z}\mid A=a,W=w,L=l\right\}\b(l\mid a,w)\g_\delta(a\mid
        w)\p(z\mid a,w)\p(w)\dd\nu(a,z,l,w)\notag\\
      &= \int \m(a,z,l,w)\b(l\mid a,w)\g_\delta(a\mid
        w)\p(z\mid a,w)\p(w)\dd\nu(a,z,l,w)\label{eq:t1e4},
  \end{align}
  where \eqref{eq:t1e1} follows by definition of $(A_\delta,G_\delta)$,
  \eqref{eq:t1e2} follows by~\ref{main:ass:ncay} and definition of
  $A_\delta$, \eqref{eq:t1e3} follows by~\ref{main:ass:ncaz}, and
  \eqref{eq:t1e4} follows by~\ref{main:ass:nczy}; the rearrangements made
  between \eqref{eq:t1e3} and \eqref{eq:t1e4} make use of~\ref{main:ass:zpos}.
  Similar arguments yield
  \[\E\{Y_{A, G}\}=\int \m(a,z,l,w)\b(l\mid a,w)\g(z\mid
    w)\p(z\mid a,w)\p(w)\dd\nu(a,z,l,w).\]
  We also have
  \begin{align*}
    \E&\{Y_{A_\delta, G}\} \notag\\&= \int \E\left\{Y_{a,z}\mid
                                     A_\delta=a,G = z,
                                     W=w\right\}\g_\delta(a\mid
                                     w)\P(G=z\mid
                                     A_\delta=a,W=w)\p(w)\dd\nu(a,z,w)\notag\\
      &= \int \E\left\{Y_{a,z}\mid W=w\right\}\g_\delta(a\mid
        w)\P(G=z\mid W=w)\p(w)\dd\nu(a,z,w)\\
      &= \int \E\left\{Y_{a,z}\mid A=a,W=w\right\}\g_\delta(a\mid
        w)\p(z\mid w)\p(w)\dd\nu(a,z,w)\\
      &= \int \E\left\{Y_{a,z}\mid A=a,W=w\right\}\g_\delta(a\mid
        w)\p(z\mid w)\p(w)\dd\nu(a,z,w)\\
      &= \int \E\left\{Y_{a,z}\mid A=a,W=w,L=l\right\}\b(l\mid a,w)\g_\delta(a\mid
        w)\p(z\mid w)\p(w)\dd\nu(a,z,l,w)\notag\\
      &= \int \m(a,z,l,w)\b(l\mid a,w)\g_\delta(a\mid
        w)\p(z\mid w)\p(w)\dd\nu(a,z,l,w).
  \end{align*}
  Subtracting gives the expressions for the PIIE and PIDE in the
  theorem.
\end{proof}

\section{Efficient influence functions (Theorem~\ref{main:theo:eif})}
\begin{proof}
  In this proof we will use $\Theta_j(\P):j=1,2$ to denote a parameter
  as a functional that maps the distribution $\P$ in the model to a
  real number. We will assume that the measure $\nu$ is discrete so
  that integrals can be written as sums, and will omit the dependence
  on $\delta$. It can be checked algebraically that the resulting
  influence function will also correspond to the influence function of
  a general measure $\nu$. The true parameter value for $\theta_1$ is
  thus given by
  \[\theta_1=\Theta_1(\P) = \sum_{y,a,z,m,w} y\,\p(y\mid a, z, l,
    w)\p(l\mid a,w)\p(z\mid a,w)\g_\delta(a\mid w)\p(w).\] The
  non-parametric MLE of $\theta_1$ in the model of $\g_\delta$ known
  is given by
  \begin{equation}
    \Theta(\Pn)=\sum_{y,a,z,m,w}y\frac{\Pn f_{y,a,z,l,w}}{\Pn
      f_{a,z,l,w}}\frac{\Pn f_{l,a,w}}{\Pn f_{a,w}}\frac{\Pn
      f_{z,a,w}}{\Pn f_{a,w}}\g_{\delta}(a\mid w)\Pn f_w\label{nonpest},
  \end{equation}
  where we remind the reader of the notation $\P f =\int f \dd\P$. Here
  $f_{y,a,z,l,w}=\one(Y=y,A=a,Z=z,M=m,W=w)$, and $\one(\cdot)$ denotes
  the indicator function. The other functions $f$ are defined
  analogously.

  We will use the fact that the efficient influence function in a
  non-parametric model corresponds with the influence curve of the
  NPMLE. This is true because the influence curve of any regular
  estimator is also a gradient, and a non-parametric model has only
  one gradient. The Delta method \cite[see, e.g., Appendix 18
  of][]{vdl2011targeted} shows that if $\hat \Theta_1(\Pn)$ is a
  substitution estimator such that $\theta_1=\hat \Theta_1(\P)$, and
  $\hat \Theta_1(\Pn)$ can be written as
  $\hat \Theta^*_1(\Pn f:f\in\mathcal{F})$ for some class of functions
  $\mathcal{F}$ and some mapping $\Theta^*_1$, the influence function of
  $\hat \Theta_1(\Pn)$ is equal to
  \[\dr_\P(O)=\sum_{f\in\mathcal{F}}\frac{\dd\hat \Theta^*_1(\P)}
  {\dd\P f}\{f(O)-\P f\}.\]

  Applying this result to (\ref{nonpest}) with
  $\mathcal{F}=\{f_{y,a,z,l,w},f_{a,z,l,w},f_{z,a,w},f_{a',w},
  f_{l,a,w},f_{a,w},f_w:y,a,z,l,w\}$ and rearranging terms gives the result of
  the theorem. The algebraic derivations involved here are lengthy and not
  particularly illuminating, and are therefore omitted from the proof. Similar
  analyses may be performed for the model where only $\g_\delta$ is unknown, as
  well as $\theta_2$.
\end{proof}

\section{Targeted minimum loss estimation algorithm}

To simplify notation, in the remaining of this section we will denote
$\tilde \eta_{j(i)}(O_i)$ with $\tilde \eta(O_i)$. If $L$ is binary,
the efficient influence functions in Theorem~\ref{main:theo:eif} may be
simplified using the following identity:
\[ \vv(l,a,w) - \bar\vv(a,w) = \{\vv(1,a,w) -\vv(0,a,w)\}
\{l-\b(1\mid a, w)\},\] which also holds for $\vv$ replaced by $\s$ and
$\bar\vv$ by $\bar\s$.

\begin{enumerate}[label=Step \arabic*., align=left, leftmargin=*]
\item Initialize $\tilde\eta =\hat\eta$. Compute $\tilde\vv$,
  $\tilde\s$, and $\tilde\q^j$ by plugging in $\tilde \m$,
  $\tilde \g$, $\tilde \e$, $\tilde \d$ into equations
  (\ref{main:eq:nuis}), (\ref{main:eq:altnuis}) and (\ref{main:eq:defqs}) if
  $Z$ is multivariate, and fitting data-adaptive regression algorithms as
  appropriate.

\item \label{step:computeH} For each subject, compute the auxiliary covariates
  \begin{align*}
    H_{\mbox{\scriptsize D}, i} &= \frac{{\tilde\b}(L_i\mid A_i,
                                  W_i)}{{\tilde\d}(L_i \mid Z_i, A_i,
                                  W_i)}\left\{1-\frac{{\tilde\g}_\delta(A_i\mid
                                  W_i)}{{\tilde\e}(A_i\mid Z_i,W_i)}\right\}\\
    H_{\mbox{\scriptsize I}, i} &= \frac{{\tilde\b}(L_i\mid A_i,
                                  W_i)}{{\tilde\d}(L_i \mid Z_i, A_i,
                                  W_i)}\left\{\frac{{\tilde\g}_\delta(A_i\mid
                                  W_i)}{{\tilde\e}(A_i\mid
                                  Z_i,W_i)}-\frac{\tilde{\g}_\delta(A_i
                                  \mid W_i)}{{\tilde\g}(A_i\mid
                                  W_i)}\right\}\\
    K_{\mbox{\scriptsize D}, i} & = {\tilde\vv}(1,A_i,W_i)
                                  - {\tilde\vv}(0,A_i,W_i) - \frac{{\tilde\g}_\delta(A_i
                                  \mid W_i)}{{\tilde\g}(A_i\mid W_i)} \{{\tilde\s}(1,A_i,W_i) -
                                  {\tilde\s}(0,A_i,W_i)\}\\
    K_{\mbox{\scriptsize I}, i} & = \frac{{\tilde\g}_\delta(A_i
                                  \mid W_i)}{{\tilde\g}(A_i\mid W_i)}
                                  \{{\tilde\s}(1,A_i,W_i) -
                                  {\tilde\s}(0,A_i,W_i) - {\tilde\vv}(1,A_i,W_i)
                                  + {\tilde\vv}(0,A_i,W_i)\}\\
    M_{\mbox{\scriptsize D}, i} & = -\frac{\tilde\g_\delta(1\mid w)(1 -
                                  \tilde\g_\delta(1\mid w))}{\tilde\g(1\mid w)(1 -
                                  \tilde\g(1\mid w))}{\tilde\q}^2(w)\\
    M_{\mbox{\scriptsize I}, i} & = \frac{\tilde\g_\delta(1\mid w)(1 -
                                  \tilde\g_\delta(1\mid w))}{\tilde\g(1\mid w)(1 -
                                  \tilde\g(1\mid w))}\{{\tilde\q}^2(w)-{\tilde\q}^1(w)\}
  \end{align*}
\item \label{step:fit} Fit the logistic tilting models
  \begin{align*}
    \logit \m_\beta(A_i,Z_i,L_i,W_i) &= \logit \tilde
                                       \m(A_i,Z_i,L_i,W_i) + \beta_I H_{\mbox{\scriptsize I}, i} +
                                       \beta_D H_{\mbox{\scriptsize D}, i}\\
    \logit \b_\alpha(1\mid A_i,W_i) &= \logit \tilde
                                      \b(1\mid A_i,W_i) + \alpha_I K_{\mbox{\scriptsize I}, i} +
                                      \alpha_D K_{\mbox{\scriptsize
                                      D}, i}\\
    \logit \g_\gamma(1\mid W_i) &= \logit \tilde
                                  \g(1\mid W_i) + \gamma_I M_{\mbox{\scriptsize I}, i} +
                                  \gamma_D M_{\mbox{\scriptsize D}, i}
  \end{align*}
  where
  $\logit(p) = \log\{p(1-p)^{-1}\}$. Here, $\logit \tilde\m(a,z,l,w)$
  is an offset variable (i.e., a variable with known parameter value
  equal to one). The parameter $\beta=(\beta_I, \beta_D)$ may be
  estimated by running standard logistic regression of $Y_i$ on
  $(H_{\mbox{\scriptsize D}, i}, H_{\mbox{\scriptsize I}, i})$ with no
  intercept and an offset term equal to
  $\logit \tilde\m(A_i,Z_i,L_i,W_i)$. Let $\hat\beta$ denote the
  estimate, and let $\tilde \m=\m_{\hat\beta}$ denote the updated
  estimates. Perform analogous computations for $\b$ and $\g$.
\item \label{step:computeu} Compute $\tilde\uu$ according to equation
  (\ref{main:eq:nuis}) by plugging in $\tilde\m$ and $\tilde\b$. Compute
  the covariate
  \[J_i = \frac{{\tilde\g}_\delta(A_i
      \mid W_i)}{{\tilde\g}(A_i\mid W_i)},\]
  and fit the model
  \[\logit \bar\uu_\kappa(A_i,W_i) = \logit \tilde{\bar\uu}(A_i,W_i) +
    \kappa_D + \kappa_I J_{ i}\] by running a logistic regression of
  $\tilde\uu(Z_i,A_i,W_i)$ on $J_i$ with an intercept and offset
  $\logit \tilde{\bar\uu}(A_i,W_i)$. Let $\hat\kappa$ denote the MLE,
  and update $\tilde{\bar\uu} = \bar\uu_{\hat \kappa}$.
\item The TMLE of the direct and indirect effects are defined as:
  \begin{equation*}
    \begin{split}
      \psidtmle &= \frac{1}{n} \int\sum_{i = 1}^n
      \left\{\tilde{\bar\uu}(a,W_i)\tilde\g(a\mid W_i) -
        \tilde\uu(Z_i,a,W_i)\tilde\g_\delta(a\mid W_i)\right\}\dd\kappa(a)\\
      \psiitmle &= \frac{1}{n} \int\sum_{i = 1}^n
      \left\{\tilde\uu(Z_i,a,W_i) -
        \tilde{\bar\uu}(a,W_i)\right\}\tilde\g_\delta(a\mid
      W_i)\dd\kappa(a)
    \end{split}
  \end{equation*}
\end{enumerate}
\section{Proof of Theorem \ref{main:theo:asos}}
\begin{proof}
  Let $\Pnj$ denote
  the empirical distribution of the prediction set ${\cal V}_j$, and let
  $\Gnj$ denote the associated empirical process
  $\sqrt{n/J}(\Pnj-\P)$. For simplicity we denote a general parameter
  $\psi$ with influence function $D_\eta$, the proof applies equally
  to the direct and indirect effect parameters. Note that
  \[\psios = \frac{1}{J}\sum_{j=1}^J\Pnj D_{\hat
      \eta_j,\delta},\,\,\,\psi_\delta=\P D_{\eta}.\]Thus,
  \[  \sqrt{n}\{\psios - \psi_\delta\}=\Gn \{D_{\eta,\delta}
    - \psi_\delta\} + R_{n,1}(\delta) + R_{n,2}(\delta),\]
  where
  \[  R_{n,1}(\delta)  =\frac{1}{\sqrt{J}}\sum_{j=1}^J\Gnj(D_{\hat
      \eta_j,\delta} - D_{\eta,\delta}),\,\,\,
    R_{n,2}(\delta)  = \frac{\sqrt{n}}{J}\sum_{j=1}^J\P\{D_{\hat
      \eta_j,\delta}-\psi_\delta\}.
  \]
  It remains to show that $R_{n,1}(\delta)$ and $R_{n,2}(\delta)$ are
  $o_P(1)$. Lemmas~\ref{main:lemma:dr1} and~\ref{main:lemma:dr2} together
  with the Cauchy-Schwartz inequality and assumption \ref{main:ass:sec1}
  of the theorem shows that $||R_{n,2}||_{\Delta}=o_P(1)$. For
  $||R_{n,1}||_{\Delta}$ we use empirical process theory to argue conditional
  on the training sample ${\cal T}_j$. In particular, Lemma 19.33
  of~\cite{vdvaart2000asymptotic} applied to the class of functions
  ${\cal F} = \{D_{\hat \eta_j,\delta} - D_{\eta,\delta}\}$ (which
  consists of one element) yields
  \[E\left\{\big|\Gnj (D_{\hat \eta_j,\delta} - D_{\eta,\delta})\big|
      \,\bigg|\, {\cal T}_j\right\}\lesssim \frac{2C\log 2}{n^{1/2}} +
    ||D_{\hat \eta_j,\delta} - D_{\eta,\delta}||(\log 2)^{1/2}\] By
  assumption \ref{main:ass:sec1}, the left hand side is $o_P(1)$. Lemma 6.1
  of~\cite{chernozhukov2018double} may now be used to argue that
  conditional convergence implies unconditional convergence, concluding
  the proof.
\end{proof}

\section{Theorem~\ref{main:theo:astmle}}
\begin{proof}
  Let $\Pnj$ denote the empirical distribution of the prediction set
  ${\cal V}_j$, and let $\Gnj$ denote the associated empirical process
  $\sqrt{n/J}(\Pnj-\P)$. For simplicity we denote a general parameter
  $\psi$ with influence function $D_\eta$, the proof applies equally
  to the direct and indirect effect parameters. By definition, the sum
  of the scores of the submodels
  $\{\m_\beta,\b_\alpha,\g_\gamma,\bar\uu_\kappa:(\beta,\alpha,
  \gamma, \kappa)\}$ at the last iteration of the TMLE procedure is
  equal to
  $n^{-1}\sum_{i=1}^n D_{\tilde \eta}(O_i) = o_P(n^{-1/2})$. Thus, we have
  \[\psitmle = \frac{1}{J}\sum_{j=1}^J\Pnj D_{\tilde \eta_j}+o_P(n^{-1/2}).\]
  Thus,
  \[  \sqrt{n}(\psitmle - \theta)=\Gn (D_{\eta}
    - \theta) + R_{n,1} + R_{n,2} + o_P(n^{-1/2}),\]
  where
  \[  R_{n,1}  =\frac{1}{\sqrt{J}}\sum_{j=1}^J\Gnj(D_{\tilde
      \eta_j} - D_{\eta}),\,\,\,
    R_{n,2}  = \frac{\sqrt{n}}{J}\sum_{j=1}^J\P(D_{\tilde
      \eta_j}-\theta).
  \]
  As in the proof of Theorem~\ref{main:theo:asos}, Lemmas~\ref{main:lemma:dr1}
  and~\ref{main:lemma:dr2} together with the Cauchy-Schwartz inequality and
  the assumptions of the theorem shows that $R_{n,2}=o_P(1)$.

  Since $D_{\tilde \eta_j}$ depends on the full sample through the
  estimates of the parameters $\beta$ of the logistic tilting models,
  the empirical process treatment of $R_{n,1}$ needs to be slightly
  from that in the proof of Theorem~\ref{main:theo:asos}. To make this
  dependence explicit, we introduce the notation
  $D_{\hat \eta_j,\beta}=D_{\tilde \eta_j}$ and $R_{n,1}(\beta)$. Let
  ${\cal F}_n^j=\{D_{\hat \eta_j,\beta} -
  D_\eta:\beta\in B\}$. Because the function
  $\hat\eta_j$ is fixed given the training data, we can apply Theorem
  2.14.2 of~\cite{vdvaart1996weak} to obtain
  \[E\left\{\sup_{f\in {\cal F}_n^j}|\Gnj f| \,\,\bigg|\,\, {\cal
        T}_j\right\}\lesssim ||F^j_n||\int_0^1\sqrt{1+N_{[\,]}(\epsilon
      ||F_n^j||, {\cal F}_n^j, L_2(\P))}\dd\epsilon, \] where
  $N_{[\,]}(\epsilon ||F_n^j||, {\cal F}_n^j, L_2(\P))$ is the
  bracketing number and we take
  $F_n^j=\sup_{\beta\in B}|D_{\hat \eta_j,\beta} -
  D_\eta|$ as an envelope for the class ${\cal
    F}_n^j$. Theorem 2.7.2 of~\cite{vdvaart1996weak} shows
  \[\log N_{[\,]}(\epsilon ||F_n^j||, {\cal F}_n^j, L_2(\P))\lesssim
    \frac{1}{\epsilon ||F_n^j||}.\]
  This shows
  \begin{align*}||F^j_n||\int_0^1\sqrt{1+N_{[\,]}(\epsilon
    ||F_n^j||, {\cal F}_n^j, L_2(\P))}\dd\epsilon &\lesssim
        \int_0^1\sqrt{||F^j_n||^2+\frac{||F^j_n||}{\epsilon}}\dd\epsilon\\
        &\leq
    ||F^j_n||+||F^j_n||^{1/2}\int_0^1\frac{1}{\epsilon^{1/2}}\dd\epsilon\\
        &\leq ||F^j_n|| + 2 ||F^j_n||^{1/2}.
  \end{align*}
  Since $||F^j_n||=o_P(1)$, this shows
  $\sup_{f\in {\cal F}_n^j}\Gnj f=o_P(1)$ for each $j$, conditional on
  ${\cal T}_j$. Thus $\sup_{\beta\in B} R_{n,1}(\beta)=o_P(1)$.
  Lemmas~\ref{main:lemma:dr1} and~\ref{main:lemma:dr2} together with the
  Cauchy-Schwartz inequality and the assumptions of the theorem show
  that $R_{n,2}=o_P(1)$, concluding the proof of the theorem.
\end{proof}

\section{Additional results}

\begin{lemma}[Second order terms for modified treatment
  policies]\label{lemma:so1}
  Let $\dd\xi(o)$ denote $\dd\nu(a,l,z)\dd\P(w)$, and let $\r(z\mid,
  a, w)$ denote $\p(z\mid a, w)$, and let $\h(z\mid,
  w)$ denote $\p(z\mid w)$. Let $d(a,w)$ denote a modified treatment
  policy satisfying~\ref{main:ass:inv}. We have
  \begin{align}
    \P D_{\eta_1,\delta}^1 -\psi_1(\delta) & =
         \int\left(\frac{\g}{\g_1}\frac{\d}{\d_1} -
        1\right)(\m-\m_1)\b_1\r\g_{\delta,1}\dd\xi\label{eq:so1}\\
        &-\int\left(\frac{\g}{\g_1}-1\right)(\bar\uu_1-\bar\uu)
        \g_{\delta,1}\dd\xi\label{eq:so3}\\
         &+\int\left(\frac{\g}{\g_1}-1\right)(\m_1-\m)\b_1\r
         \g_{\delta,1}\dd\xi\label{eq:so4}\\
         &-\int\frac{\g}{\g_1}(\b_1-\b)(\vv_1-\vv)\g_{\delta,1}
         \dd\xi\label{eq:so5}\\
        &-\int(\bar\uu_1-\bar\uu)(\g_{\delta,1}-\g_\delta)
        \dd\xi\label{eq:so2}
  \end{align}
  and
  \begin{align*}
    \P D_{\eta_1,\delta}^2 - \psi_2(\delta)
    & = \int\left(\frac{\e}{\e_1}\frac{\d}{\d_1} -
      1\right)(\m-\m_1)\b_1\h\g_{\delta,1} \dd\xi\\
    &+\int\frac{\g}{\g_1}(\b_1-\b)(\s_1-\s)\g_{\delta,1}\dd\xi\\
    &-\int(\q_1-\q)(\g_{\delta,1}-\g_\delta)\dd\xi.
  \end{align*}
\end{lemma}
\begin{proof} Note that
\begin{equation}
  \begin{split}
    \P S_{\eta_1,\delta}^1 -
    \psi_1(\delta)
    & =
      \int\left(\frac{\g}{\g_1}\frac{\d}{\d_1}
      - 1\right)(\m-\m_1)\b_1\r\g_{\delta,1}\dd\xi
      + \int(\m-\m_1)\b_1\r\g_{\delta,1}\dd\xi                                  \\
    & - \int\bar\uu
      (\g_\delta - \g_{\delta,1})\dd\xi - \int \bar\uu
      \g_{\delta,1}\dd\xi                                                       \\
    & +\int
      \frac{\g}{\g_1}(\b-\b_1)\vv_1\g_{\delta,1}\dd\xi
      + \int
      \frac{\g}{\g_1}(\uu_1\r-\bar\uu_1)\g_{\delta,1}\dd\xi
      +\int\bar\uu_1\g_{\delta,1}\dd\xi                                         \\
    & =(\ref{eq:so1})
      +\int (\bar\uu_1-\bar\uu)\g_{\delta,1}\dd\xi                              \\
    & +\int(\m-\m_1)\b_1\r\g_{\delta,1}\dd\xi +
      \int\frac{\g}{\g_1}(\b-\b_1)\vv_1\g_{\delta,1}\dd\xi                      \\
    & + \int
      \frac{\g}{\g_1}\uu_1\r\g_{\delta,1}\dd\xi
      -\int
      \frac{\g}{\g_1}\bar\uu_1\g_{\delta,1}\dd\xi                               \\
    & - \int\bar\uu
      (\g_\delta - \g_{\delta,1})\dd\xi           \\
    & =(\ref{eq:so1})
      -\int\left(\frac{\g}{\g_1}-1\right)(\bar\uu_1-\bar\uu)\g_{\delta,1}\dd\xi \\
    & +\int \frac{\g}{\g_1}\m_1\b_1\r\g_{\delta,1}\dd\xi - \int
      \frac{\g}{\g_1}\m\b\r\g_{\delta,1}\dd\xi                                  \\
    & +\int(\m-\m_1)\b_1\r\g_{\delta,1}\dd\xi +
      \int\frac{\g}{\g_1}(\b-\b_1)\vv_1\g_{\delta,1}\dd\xi                      \\
    & - \int\bar\uu
      (\g_\delta - \g_{\delta,1})\dd\xi           \\
    & =(\ref{eq:so1})-
      (\ref{eq:so3}) \\
    & +\int(\m-\m_1)\b_1\r\g_{\delta,1}\dd\xi +
      \int\frac{\g}{\g_1}(\b-\b_1)\vv_1\g_{\delta,1}\dd\xi            \\
    & +\int
      \frac{\g}{\g_1}(\m_1\b_1
      +\m\b_1-\m\b_1-
      \m\b)\r\g_{\delta,1}\dd\xi                                      \\
    & - \int\bar\uu
      (\g_\delta - \g_{\delta,1})\dd\xi \\
    & =(\ref{eq:so1}) - (\ref{eq:so3})                                                  \\
    & +
      \int(\m-\m_1)\b_1\r\g_{\delta,1}\dd\xi +
      \int\frac{\g}{\g_1}(\b-\b_1)\vv_1\g_{\delta,1}\dd\xi            \\
    & +\int
      \frac{\g}{\g_1}(\m_1
      -\m)\b_1\r\g_{\delta,1}\dd\xi
      +\int
      \frac{\g}{\g_1}(\b_1-\b)\m\r\g_{\delta,1}\dd\xi                 \\
    & - \int\bar\uu
      (\g_\delta - \g_{\delta,1})\dd\xi \\
    & =(\ref{eq:so1})-(\ref{eq:so3}) +(\ref{eq:so4})-(\ref{eq:so5}) \\
    & - \int\bar\uu(\g_\delta - \g_{\delta,1})\dd\xi.
  \end{split}\label{eq:proof1}
\end{equation}
Using~\ref{main:ass:inv} we can change variables to obtain
\[\P S_{\eta_1,\delta}^{A,1} =
  \int\bar\uu_1(\g_\delta-\g_{\delta,1})\dd\xi.\] The proof for
$\psi_2$ is analogous. This completes the
proof of the theorem.
\end{proof}

\begin{lemma}[Second order terms for
  exponential tilting.]\label{lemma:so2}
  Define $c(w) = \{\int_a \exp(\delta a)\g(a\mid w)\}^{-1}$, and let
  $c_1(w)$ be defined analogously. Let $b(a) = \exp(\delta a)$. Using
  the same notation as in Lemma~\ref{main:lemma:dr1}, we have
  \begin{align*}
    \P D_{\eta_1,\delta}^1  -\psi_1(\delta)
    & =
      \int\left(\frac{\g}{\g_1}\frac{\d}{\d_1} -
      1\right)(\m-\m_1)\b_1\r\g_{\delta,1}\dd\xi\\
    &-\int\left(\frac{\g}{\g_1}-1\right)(\bar\uu_1-\bar\uu)\g_{\delta,1}\dd\xi\\
    &+\int\left(\frac{\g}{\g_1}-1\right)(\m_1-\m)\b_1\r\g_{\delta,1}\dd\xi\\
    &-\int\frac{\g}{\g_1}(\b_1-\b)(\vv_1-\vv)\g_{\delta,1}\dd\xi\\
    &+ \int(\bar\uu_1-\bar\uu)(\g_{\delta,1}-\g_\delta)\dd\xi\\
    & -\int\left\{(c_1-c)^2\int b
      \g_1\bar\uu_1\dd\kappa\int b
      \g\dd\kappa\right\}\dd\xi\\
    &+\int \left\{(c_1-c)\int b\bar\uu_1(\g-\g_1)\dd\kappa\right\}\dd\xi,
  \end{align*}
  and
  \begin{align*}
    \P D_{\eta_1,\delta}^2  -\psi_2(\delta)
    & = \int\left(\frac{\e}{\e_1}\frac{\d}{\d_1}  -
      1\right)(\m-\m_1)\b_1\h\g_{\delta,1} \dd\xi\\
    &+\int\frac{\g}{\g_1}(\b_1-\b)(\s_1-\s)\g_{\delta,1}\dd\xi\\
    &-\int(\q_1-\q)(\g_{\delta,1}-\g_\delta)\dd\xi\\
    & -\int\left\{(c_1-c)^2\int b
      \g_1\bar\q_1\dd\kappa\int b
      \g\dd\kappa\right\}\dd\xi\\
    &+\int \left\{(c_1-c)\int b\bar\q_1(\g-\g_1)\dd\kappa\right\}\dd\xi.
  \end{align*}

\end{lemma}

\begin{proof}
  In this proof, (\ref{eq:proof1}) is also valid. We have
  \[\P S_{\eta_1,\delta}^{1,A} - \int\bar\uu(\g_\delta -
    \g_{\delta,1})\dd\xi = \P S_{\eta_1,\delta}^{1,A} - \int\bar\uu_1(\g_\delta -
    \g_{\delta,1})\dd\xi + \int(\bar\uu_1-\bar\uu)(\g_{\delta,1}-\g_\delta)\dd\xi\]
It thus remains to
  prove that
  \begin{align*}
    \P S_{\eta_1,\delta}^{1,A} - \int\bar\uu_1(\g_\delta -
    \g_{\delta,1})\dd\xi = & -\int\left\{(c_1-c)^2\int b
      \g_1\bar\uu_1\dd\kappa\int b
      \g\dd\kappa\right\}\dd\xi\\
    &+\int \left\{(c_1-c)\int b\bar\uu_1(\g-\g_1)\dd\kappa\right\}\dd\xi.
  \end{align*}
  We have
  \begin{align}
    \P S_{\eta_1}^{1,A} -
    & \int\bar\uu_1(\g_{\delta}-\g_{\delta,1})\dd\xi\notag\\
    &=\int\left\{\int\frac{\g_{1,\delta}}{\g_1}\bar\uu_1 \g\dd\kappa
      -\int \frac{\g_{1,\delta}}{\g_1}\g\dd\kappa\int\bar\uu_1 \g_{1,\delta}\dd\kappa + \int(\g_{1,\delta} -
      \g_\delta)\bar\uu_1\dd\kappa\right\}\dd\xi\notag\\
    &=\int\left\{\frac{\g_{1,\delta}}{\g_1}\g\bar\uu_1\dd\kappa - \int \g_\delta\bar\uu_1\dd\kappa + \int
      \g_{1,\delta}\bar\uu_1\dd\kappa\left[1-\int\frac{\g_{1,\delta}}{\g_1}\g\dd\kappa\right]\right\}
      \dd\xi\notag\\
    &=\int\left\{c_1\int
      b\bar\uu_1 \g\dd\kappa
      - c_1\int
      b\bar\uu_1
      \g\dd\kappa
      + c_1\int
      b\g\bar\uu_1\dd\kappa\int(c-c_1)b\g\dd\kappa\right\}\dd\xi\notag\\
    &=\int(c_1-c)\left\{\int b\bar\uu_1 \g\dd\kappa - c_1\int b
      \g_1\bar\uu_1\dd\kappa\int b \g\dd\kappa\right\}\dd\xi\notag\\
    &=\int(c_1-c)\left\{\int b\bar\uu_1 \g\dd\kappa - c\int b
      \g_1\bar\uu_1\dd\kappa\int b \g\dd\kappa - (c_1-c)\int b
      \g_1\bar\uu_1\dd\kappa\int b \g\dd\kappa\right\}\dd\xi\notag\\
    &=\int\left\{-(c_1-c)^2\int b
      \g_1\bar\uu_1\dd\kappa\int b \g\dd\kappa + (c_1-c)\left[\int
      b\bar\uu_1 \g\dd\kappa - \int
      b\g_1\bar\uu_1\dd\kappa\right]\right\}\dd\xi\label{eq:intone}\\
    &=\int\left\{-(c_1-c)^2\int b
      \g_1\bar\uu_1\dd\kappa\int b \g\dd\kappa + (c_1-c)\int b\bar\uu_1(\g-\g_1)\dd\kappa\right\}\dd\xi\notag,
  \end{align}
  where (\ref{eq:intone}) follows from $c\int b \g\dd\kappa = 1$. The proof for
  $\psi_2$ is analogous.
\end{proof}

\bibliographystyle{plainnat}
\bibliography{refs}

%% file: _preamble.tex
\usepackage{float}
\usepackage{graphicx}
\usepackage{comment}
\graphicspath{ {figs/} }
\usepackage[english]{babel}
\usepackage{dsfont}
\usepackage[OT1]{fontenc}
\usepackage[utf8]{inputenc}
\usepackage{subcaption}
\usepackage{refcount}
\usepackage{lmodern}
\usepackage{color}
\usepackage{booktabs}
\usepackage{natbib}
\usepackage{tikz}
\usepackage{hyperref}

\usepackage{nameref,zref-xr} 
\zxrsetup{toltxlabel} 
\usepackage{url}

\usepackage{mathtools}
\usepackage{amsfonts}
\usepackage{amssymb}
\usepackage{bbm}
\usepackage{amsthm}
\usepackage{amsxtra}
\usepackage{amsmath}
\usepackage{commath}
\usepackage{times}
\usepackage{bm}

\usepackage{cleveref}

\let\tiny\normalsize
\let\scriptsize\normalsize
\let\footnotesize\normalsize
\let\small\normalsize
\let\large\normalsize
\let\Large\normalsize
\let\LARGE\normalsize
\let\huge\normalsize
\let\Huge\normalsize
\usepackage[inline,shortlabels]{enumitem}

\newcommand{\titlepaper}{Nonparametric causal mediation analysis for
  stochastic interventional (in)direct effects}

\newcommand{\authorlistarxiv}{
    Nima S.~Hejazi \\
    Division of Biostatistics, \\
    Department of Population Health Sciences, \\
    Weill Cornell Medicine \\
    \texttt{nhejazi@berkeley.edu} \\
    \And
    Kara E.~Rudolph\\
    Department of Epidemiology, \\
    Mailman School of Public Heatlh, \\
    Columbia University \\
    \texttt{kr2854@cumc.columbia.edu} \\
    \And
    Mark J.~{van der Laan} \\
    Division of Biostatistics, \\
    School of Public Health, and\\
    Department of Statistics,\\
      University of California, Berkeley \\
    \texttt{laan@berkeley.edu} \\
    \And
    Iv\'an D\'iaz \\
    Division of Biostatistics, \\
    Department of Population Health Sciences, \\
    Weill Cornell Medicine \\
    \texttt{ild2005@med.cornell.edu} \\
}

{\theoremstyle{lemma} }

\newcommand{\supp}{\mathop{\mathrm{supp}}}
{\theoremstyle{definition} \newtheorem{assumptioniden}{}}

\AtEndDocument{\refstepcounter{theorem}\label{finalthm}}
\AtEndDocument{\refstepcounter{equation}\label{finaleq}}

\DeclareMathOperator{\logit}{logit}
\DeclareMathOperator{\var}{Var}

\renewcommand{\P}{\mathsf{P}}
\newcommand{\m}{\mathsf{m}}
\newcommand{\p}{\mathsf{p}}
\newcommand{\q}{\mathsf{q}}

\renewcommand{\b}{\mathsf{b}}
\renewcommand{\d}{\mathsf{d}}
\newcommand{\g}{\mathsf{g}}

\newcommand{\e}{\mathsf{e}}
\newcommand{\uu}{\mathsf{u}}
\newcommand{\vv}{\mathsf{v}}
\newcommand{\s}{\mathsf{s}}
\newcommand{\indep}{\mbox{$\perp\!\!\!\perp$}}

\newcommand{\dd}{\mathrm{d}}

\newcommand{\Pn}{\mathsf{P}_n}

\newcommand{\psii}{\psi_{I, \delta}}
\newcommand{\psid}{\psi_{D, \delta}}

\newcommand{\psidos}{\hat\psi_{D, \delta}^{os}}
\newcommand{\psiios}{\hat\psi_{I, \delta}^{os}}
\newcommand{\psidtmle}{\hat\psi_{D, \delta}^{tmle}}
\newcommand{\psiitmle}{\hat\psi_{I, \delta}^{tmle}}

\newcommand{\E}{\mathsf{E}}
\newcommand{\M}{\mathcal{M}}

\pgfdeclarelayer{background}
\pgfsetlayers{background,main}
\usetikzlibrary{arrows,positioning}
\tikzset{
>=stealth',
punkt/.style={
rectangle,
rounded corners,
draw=black, very thick,
text width=6.5em,
minimum height=2em,
text centered},
pil/.style={
->,
thick,
shorten <=2pt,
shorten >=2pt,}
}
\newcommand{\Vertex}[2]
{\node[minimum width=0.6cm,inner sep=0.05cm] (#2) at (#1) {$\footnotesize#2$};
}
\newcommand{\Vertexr}[2]
{\node[rectangle, draw, minimum width=0.6cm,inner sep=0.05cm] (#2) at (#1) {$\footnotesize#2$};
}
\newcommand{\ArrowR}[3]%
{ \begin{pgfonlayer}{background}
\draw[->,#3] (#1) to[bend right=30] (#2);
\end{pgfonlayer}
}
\newcommand{\ArrowL}[3]%
{ \begin{pgfonlayer}{background}
\draw[->,#3] (#1) to[bend left=45] (#2);
\end{pgfonlayer}
}
\newcommand{\EdgeL}[3]%
{ \begin{pgfonlayer}{background}
\draw[dashed,#3] (#1) to[bend right=-45] (#2);
\end{pgfonlayer}
}
\newcommand{\Arrow}[3]%
{ \begin{pgfonlayer}{background}
\draw[->,#3] (#1) -- +(#2);
\end{pgfonlayer}
}